\documentclass{aa}
\usepackage{txfonts}
\usepackage{graphics}
\usepackage{epsfig}
\newcommand{\jk}{\mbox{$J\!-\!K$}}

\newcommand{\mh}{\mbox{\rm [{\rm M}/{\rm H}]}}

\newcommand{\Teff}{\mbox{$T_{\rm eff}$}}

\newcommand{\comment}[1]{}

\begin{document}

   \title{AGB stars in the Magellanic Clouds. III. The rate of star
   formation across the SMC.}

   \author{M.-R.L. Cioni\inst{1}
          \and
           L. Girardi\inst{2}
          \and
           P. Marigo\inst{3}
           \and
           H.J. Habing\inst{4}
          }

   \offprints{mrc@roe.ac.uk}

   \institute{SUPA, School of Physics, University of Edinburgh,
              IfA, Blackford Hill, Edinburgh EH9 3HJ, UK
              \and
              Osservatorio Astronomico di Padova, INAF, 
              Vicolo dell'Osservatorio 5, 35122 Padova, Italy
              \and
              Dipartimento di Astronomia, Universit\`{a} di Padova,
              Vicolo dell'Osservatorio 2, 35122 Padova, Italy
              \and
              Sterrewacht Leiden,
              Niels Bohrweg 2, 2333 RA Leiden, The Netherlands
             }

   \date{Received 16 December 2005 / Accepted 21 February 2006}

   \titlerunning{SFR across the SMC}

   \authorrunning{Cioni et al.}

   \abstract{}{This article compares the $K_\mathrm{s}$ magnitude
distribution of Small Magellanic Cloud asymptotic giant branch stars
obtained from the DENIS and 2MASS data with theoretical
distributions.}{Theoretical $K_\mathrm{s}$ 
     magnitude distributions have been constructed using
up-to-date    stellar    evolution    calculations   for    low    and
intermediate-mass stars,  and in particular for  thermally pulsing
asymptotic  giant  branch   
stars.  Separate fits of the magnitude distributions of carbon-  and
oxygen-rich stars  allowed   us  to  constrain  the  metallicity
distribution across the galaxy and its star formation rate.}{The Small
     Magellanic Cloud 
stellar population is found to be on average $7-9$ Gyr old but older
stars are present at its
periphery and younger stars are  present in the direction of the
companion galaxy the Large  Magellanic  Cloud.    
The  metallicity  distribution  traces a 
ring-like structure that  is more metal rich than  the inner region of
the galaxy.} 
{The C/M ratio  discussed in Paper I is a tracer of the metallicity
distribution only if the underlying stellar population is of
intermediate-age.}

\maketitle

\section{Introduction}
Asymptotic Giant  Branch (AGB) stars are useful indicators of
galactic structure and evolution for a series of reasons:
first, they are easily noticed in near-infrared surveys, 
second, they trace stellar populations over a large range of 
ages (from $\sim0.1$ to several Gyr) and hence are widely 
distributed among galaxies, and third, their distributions of 
pulsational and chemical properties are sensitive to population 
parameters such as age and metallicity.
Cioni \& Habing (\cite{crat}) -- 
herewith Paper I -- showed that the number ratio between carbon-rich 
(C-rich or C-type) and oxygen-rich (O-rich or M-type) AGB stars varies 
over the face of the Magellanic Clouds.  Interpreting C/O changes as 
changes in the mean metallicity, they 
concluded that in the Small Magellanic Cloud 
(SMC) contrary to the Large Magellanic Cloud (LMC) there is no clear gradient 
in metallicity. The results for the LMC have been revised by 
Cioni et al. (\cite{cio2}) -- herewith Paper II --
with the aid of up-to-date stellar models and taking into
account different age distributions. It was suggested that a 
fit to the $K_{\mathrm s}$-band magnitude distribution of both 
C- and O-rich AGB stars should be particularly useful to detect 
variations on the mean age and metallicity across the face of 
nearby galaxies. For some well-defined LMC regions, both the 
mean age and metallicity were found to span the whole range of 
grid parameters. 

This work, similar to Paper II, focuses on the stellar population in 
the SMC field. The SMC is also an interesting test study to our
method based on the near-infrared photometry, since the global star 
formation history in this galaxy has been recently studied by 
Harris \& Zaritsky (\cite{harr}) using photometry in the optical.
Below, we provide a summary of previous work on the star formation 
history of the SMC.

Section \ref{obs} describes our selection of the sample of AGB
stars from the DENIS catalogue towards the Magellanic Clouds
 and from the 2MASS catalogue. Section \ref{theo} summarises 
the theoretical models used  to construct a magnitude distribution 
%(a detailed description of the models is given in Paper II)
while Sect. \ref{comp} compares the observed and theoretical
distributions. Section \ref{dis} contains a comparison of our
results with the information known from the literature and 
Section \ref{fin} concludes this paper.

\subsection{SFR in the SMC - Review}
\label{intro}

\begin{figure}
\resizebox{\hsize}{!}{\includegraphics{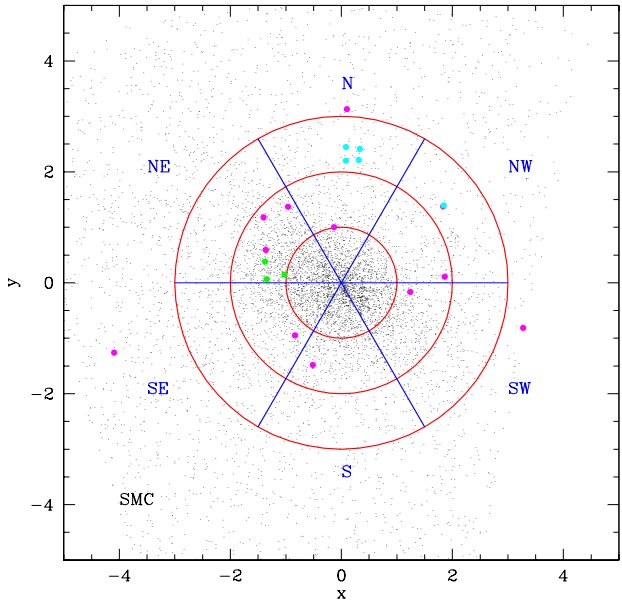}}
\caption{Distribution of the AGB stars  in the SMC.  Data are from the
DCMC and include only sources detected in three DENIS wave bands.  The
division into rings and sectors used in this article is indicated. The
centre is at $\alpha = 12.5^{\circ}$ and $\delta = -73^{\circ}$.
Thick dots indicate the location of
regions analysed by other authors, in particular: Hardy \& Durand
(\cite{hardur}, green), Crowl et al. (\cite{cro}, magenta) and Dolphin et
al. (\cite{dol}, cyan).}
\label{smcagb}
\end{figure}

The SMC, like its companion LMC galaxy, is also characterised by a 
bar, that is
probably of tidal origin, with  a  position angle  of $45^{\circ}$
and  a wing  situated  about $2^{\circ}$ East of  the bar about
$3$ kpc closer to  us than the main body of the galaxy (van den
Bergh \cite{vdbe}). The SMC is $6-12$ kpc thick along the line of
sight (Crowl et al. \cite{cro}) and appears as a smooth elliptical galaxy
from star counts of giant stars compared to the  irregular shape
traced  by younger  components (i.e.  Rebeirot et al. \cite{reb},
Morgan  \&  Hatzidimitriou \cite{morhat},  Cioni  et al.
\cite{cmor}). This galaxy, compared to the LMC, has a lower
metallicity, a higher gas mass fraction, and the mean age of its 
field stellar population is older.
It contains only one globular cluster (NGC 121; $12\pm2$ Gyr old). 
All these are indications that the SMC is more primitive and less 
evolved than the LMC (van den Bergh \cite{vdbe}). The birth rate
of clusters in the SMC has been rather constant in time
(i.e. Frantsman \cite{fra}).  Two peaks
in the  age distribution of SMC clusters  have been found: one
$6.5$ Gyr ago and the other $2.5$ Gyr ago. Both are associated
with a close passage of the LMC (Piatti et al. \cite{pia}). 
Despite the apparent abundance of young ($<3.5$ Gyr) clusters Rafelski 
\& Zaritsky (\cite{rafzar}) suggest that the dominant epoch of cluster 
formation was the initial one. Prior to their work which include about $200$ 
clusters across the SMC the distribution of young clusters
was biased towards  the East (Crowl et al. \cite{cro}, see
our Fig. \ref{smcagb}). However, because of the limited precision in the age 
measurements using integrated colours Rafelski \& Zaritsky (\cite{rafzar}) 
fail to establish a correspondence between the cluster age function and the 
field star formation history. But their Fig. $15$ suggests that clusters older 
than $3.5$ Gyr occupy the SMC  bar region while younger clusters define a 
thicker bar almost tracing the whole extent of the galaxy. 

Hardy \& Durand (\cite{hardur}) studied the ($B-V$, $B$)
colour-magnitude diagram of several fields in the wing of the SMC
(Fig. \ref{smcagb}) and concluded that the underlying population
is  of intermediate-age (older than $3$ Gyr). On the other hand
from the analysis of blue and red UK Schmidt Telescope
photographic plates, covering approximately the whole galaxy,
Gardiner   \& Hatzidimitriou (\cite{garhat}) found that  a
population  younger than $2$ Gyr is more prominent in the eastern 
side of the SMC facing the LMC than elsewhere in the galaxy, a
population $3-4$ Gyr old is present in the innermost region  of
the galaxy, while an  older population (about $10$ Gyr old)
dominates the outer regions. Dolphin et al. (\cite{dol}) observed the
outer SMC (see our Fig. \ref{smcagb}) using the Hubble Space
Telescope (HST) and derived a wide range of ages (from $2$ to
$9-12$  Gyr old) for its stellar population. Large scale $UBVI$
photometric observations by Harris \&  Zaritsky (\cite{harr})
provided for the first  time a picture of  the global star
formation history across the SMC. 
They have shown that about half of the stars in
the SMC formed more than $8.4$ Gyr ago; star formation then became
quiet until about $3$ Gyr ago after which several bursts of star formation, 
superimposed on a continuous process extending to the present time, occurred
at look-back times of $2.5$, $0.4$ and $0.06$ Gyr. The first two
bursts are associated with tidal interactions with the Milky Way
Galaxy and the latter with an interaction with the  LMC (Zaritsky \&
Harris \cite{zarhar}). Harris \& Zaritsky (\cite{harr}) also detected a 
ring-like morphology in the intermediate-age population ($1-2.5$ Gyr) that may 
suggest an inward propagation of star formation or the remnant of a gas-rich 
merger event. 

\vspace{0.2cm}
Thus, apart from the quite comprehensive study by 
Harris \& Zaritsky (\cite{harr}), based on relatively shallow photometry,
there have been considerably fewer (and less detailed) 
observations of the SMC
field stellar population and clusters than of its more proximate
companion (i.e. the LMC). Both clusters and field stars forming times
span a broad and continuous range of ages. The underlying stellar
population is also of intermediate age but there is a much older
population in the outer regions and perhaps a younger component in the
wing region.  The global picture provided by Harris \& Zaritsky (\cite{harr})
provides an extremely useful reference to future studies of the 
SMC population, including the present work.

\section{Observational sample of AGB stars}
\label{obs}

Following the same criteria as in Paper I \& II we selected an almost
complete sample of AGB  stars. From the DENIS catalogue towards the
Magellanic Clouds (DCMC - Cioni et al. \cite{ccat}) we extracted $7652$ 
candidate AGB stars including O-rich AGB stars of early M spectral subtype 
(i.e. M0-M1) but excluding AGB stars with thick circumstellar envelopes. 
First, we dereddened the data adopting an average $E(B-V)=0.065$ as given by 
Westerlund (\cite{west}), which using the extinction law by Glass 
(\cite{glas}) results in the following absorptions: $A_I=0.11$, $A_J=0.05$ and 
$A_{K_{\mathrm s}}=0.01$ mag. 
Then, we adjusted the data to the 2MASS photometry using 
the systematic shifts derived by Delmotte et al. (\cite{delm}). Finally, we 
selected AGB stars using the $(I_0, (I-J^{2MASS})_0)$ colour-magnitude diagram 
(see Fig. 1 of Cioni, Habing \& Israel \cite{cmor} and Paper II). Besides 
accounting for the different distance moduli of the Clouds 
(i.e. $\delta(m-M)_0=-0.41$ mag) no correction was applied to
compensate for the different average 
metallicity (i.e. [Fe/H]$_{LMC}=-0.30$ and [Fe/H]$_{SMC}=-0.73$ -- van den 
Berg \cite{vdbe}). The effect of metallicity on the $(I-J)$ colour is 
comparable to the photometric errors at the AGB magnitudes (i.e. 
$\sigma_I<0.02$ and $\sigma_{I-J}<0.03$ mag). 

From the 2MASS all sky  survey (Skrutskie \cite{skru}) 
we extracted $8567$ candidate AGB stars. However, 
in this case selecting them from the $((J-K_{\mathrm s})_0,
K_{{\mathrm s},0})$  
diagram we accounted for both differences in distance and metallicity
between the LMC and SMC. 
The latter causes a significant
effect on the ($J-K_s$) colour of red giants: $\Delta(J-K_s)  \simeq
0.118\times \Delta{\mathrm {[Fe/H]}}=0.05$ mag where the $0.118$ factor
is derived from our    theoretical simulations    at different
metallicities. 

To  study  the shape  of  the magnitude  distributions as  a
function of spatial coordinates  we divided the SMC area into
rings and sectors (Fig. \ref{smcagb}): three concentric rings
centred at $\alpha = 12.5^{\circ}$  and $\delta = -73^{\circ}$ at a
radius of $1^{\circ}$,  $2^{\circ}$ and $3^{\circ}$, respectively, as
well as six sectors  at an aperture  of $60^{\circ}$. Rings  are numbered
$0-2$ with increasing radius and sectors by compass direction (SE, S, SW,
NW, N, NE). 

In each sector of each ring simple averages of DCMC and 2MASS
distributions of C and M stars as a function of $K_{\mathrm s}$
magnitude have been calculated for magnitudes above the TRGB ($K_{\mathrm
s}<12.6$). Irrespective of their $I$ or $H$ magnitudes both data sets
should have detected the same AGB star candidates, and the averaging
accounts for migration between adjacent bins and variability effects. 
At $K_{\mathrm s}>12.6$ only DCMC data provide a reliable description
of the AGB population because the initial selection criterion using $I$-band
measurements rejects RGB stars with similar infrared magnitudes and
colours. More details about the comparison between DCMC and 2MASS data
are given in Paper II.

\section{Theoretical models}
\label{theo}

\begin{figure}
\resizebox{0.9\hsize}{!}{\includegraphics{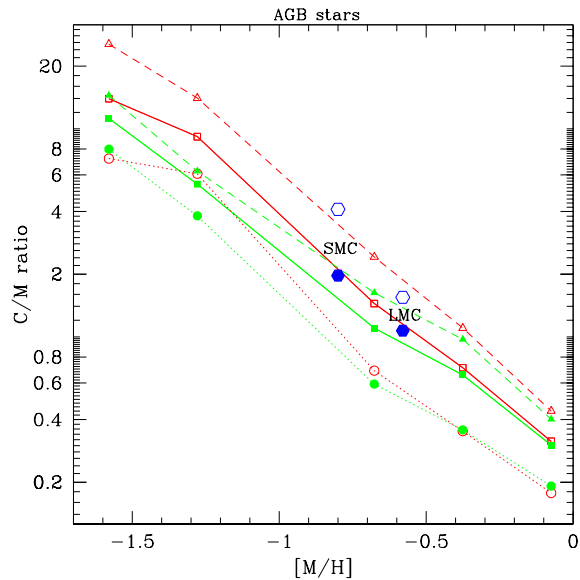}}
\caption{Variation of C/M with $\mh$ for three different cases of
SFR -- from top to bottom: $\propto\exp(-t/5\,{\rm Gyr})$ (triangles
and dashed lines), constant (squares and continuous lines), and
$\propto\exp(t/5\,{\rm Gyr})$ -- and using both the model (empty
symbols) and photometric (filled symbols) selection criteria for
AGB stars above the TRGB. The hexagons represent the C/M for 
``realistic'' models of the LMC (red; see Paper~II) and SMC (green) 
populations.} 
\label{fig_CM_SFR}
\end{figure}

\begin{figure}
\resizebox{\hsize}{!}{\includegraphics{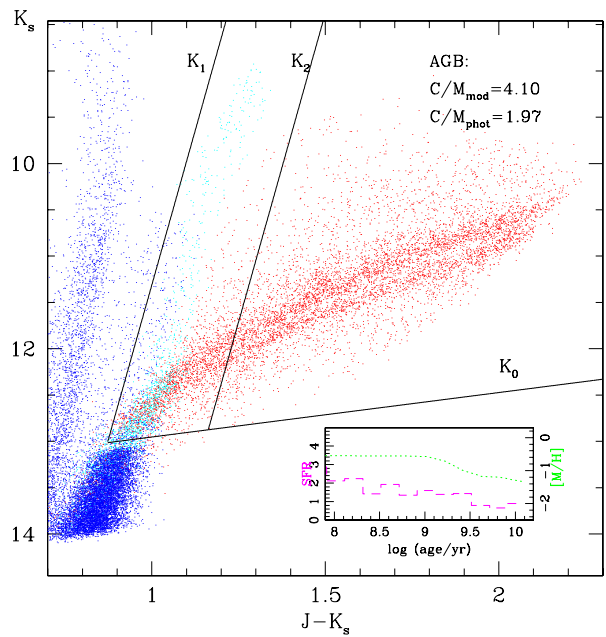}}
\caption{ Simulated CMDs for a model of the entire SMC using
Harris \& Zaritsky's (\cite{harr}) SFR, and
Pagel \& Tautvaisiene's (\cite{page}) AMR. The simulation is complete
down to $K_{\rm s}\simeq 14$, and the TRGB can be easily identified
at $K_{\rm s}\simeq 13$.
In the electronic version of this paper only, different kinds of stars
are denoted with different colours, namely blue for all stars 
before the TP-AGB, cyan for O-rich and red for C-rich TP-AGB stars.
%this simulation the O-rich TP-AGB stars make a single and
%straight sequence in the CMD, being much better defined than in
%constant-metallicity simulations. 
} 
\label{fig_compMCs}
\end{figure}

\begin{figure*}
\begin{minipage}{0.45\textwidth}
\resizebox{\hsize}{!}{\includegraphics{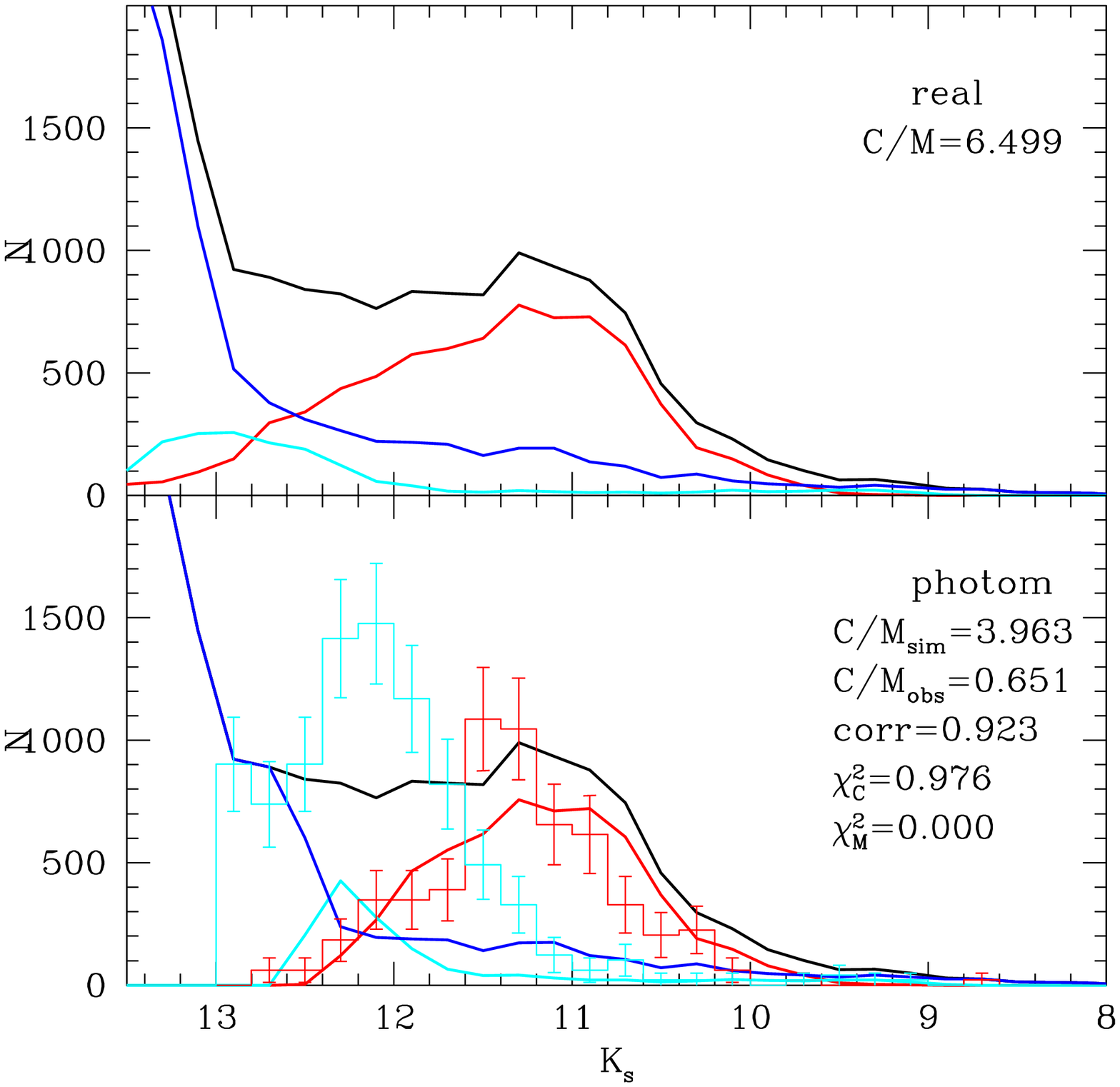}}
\end{minipage}
\hfill
\begin{minipage}{0.45\textwidth}
\resizebox{\hsize}{!}{\includegraphics{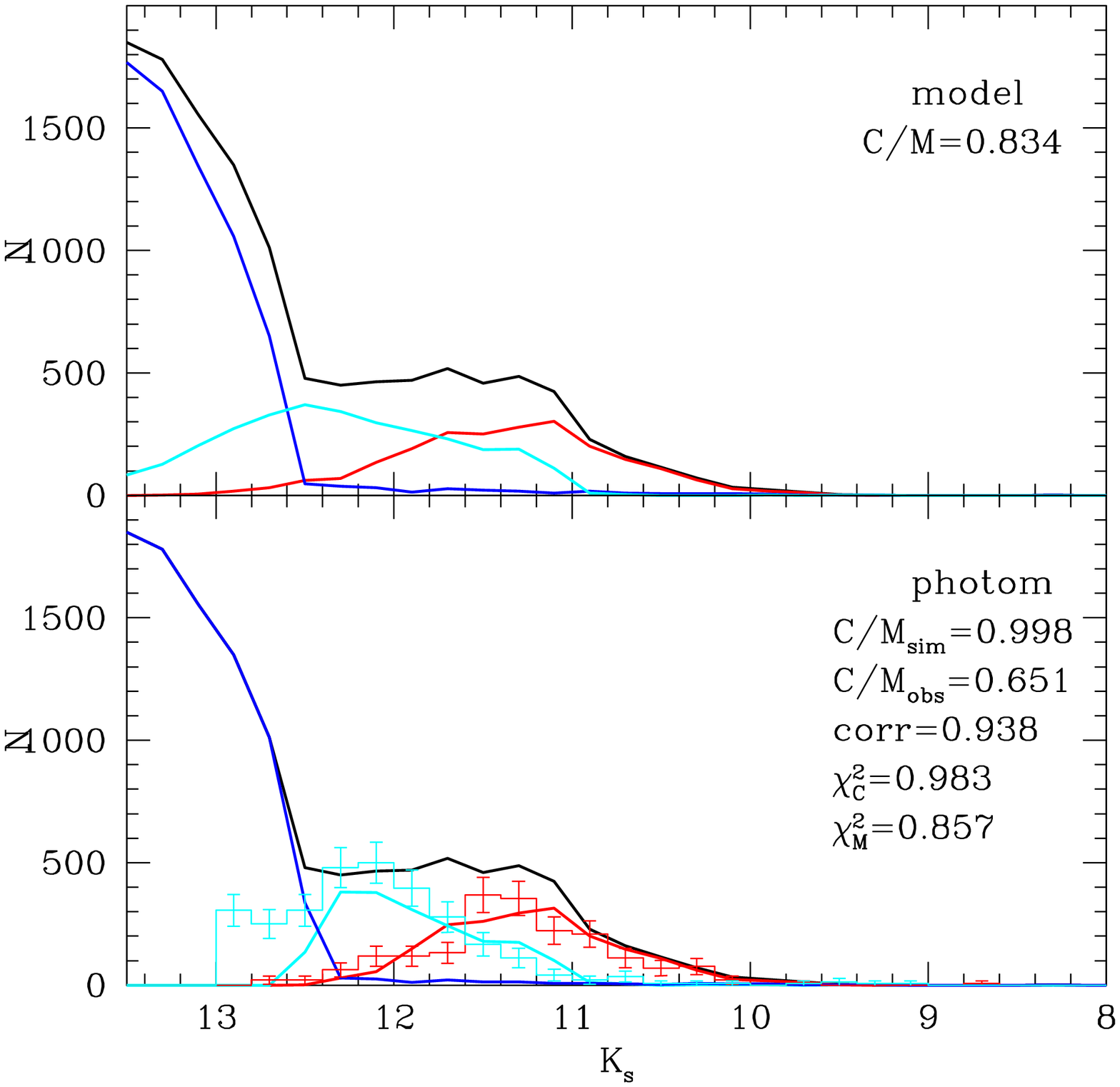}}
\end{minipage}
\caption{ LFs for a ``realistic'' SMC model ({\bf left}) and 
for a model that fits rather well both the
distribution of C and M stars ({\bf right}) in sector N of ring 0. 
% Z=0.004 and SFR=5, ring e0
The continuous lines show model results: cyan =
O-rich TP-AGB, red = C-rich TP-AGB, blue = other stars, black =
total. Upper panel: using the modelling criteria to classify
TP-AGB stars; bottom panel: using the photometric criteria. In the
latter
 case, LFs are compared to LMC and SMC data (histograms).
The number counts in the simulations have been normalised to
provide the same C-star number as in the data. In each panel of
the bottom row the labels inside each figure give  the observed
and photometric C/M ratios as well as the correlation coefficient
between the theoretical and observed distribution of C stars. The
last two numbers indicate the probability that the observed
distribution is represented by the given model distribution;  this
is the same probability to obtain the $\chi^2$ derived from the
comparison between the two distributions.} \label{fig_LFs}
\end{figure*}

The $JK_s$-band photometry of the SMC has been
simulated using the TRILEGAL population  synthesis   code  
(Girardi  et al. \cite{gi05}), that randomly generates a population
of stars following a given SFR,
age-metallicity relation (AMR) and initial  mass   function (IMF).  
The   stellar  intrinsic  properties 
are interpolated over  a large grid of stellar evolutionary
tracks,  based on  Bertelli et  al. (\cite{bert}) 
and Girardi et al. (2003) for  massive
stars, Girardi et al. (\cite{gi00}) for low- and intermediate-mass
stars, and complemented with grids of thermally pulsing AGB (TP-AGB)
tracks calculated by 
means of Marigo et al.'s (\cite{ma99}) synthetic code. Improvements to
the code are described in detail in Marigo et al. (\cite{ma03})
and in Paper II. 
We use complete grids of tracks for $5$
different metallicities comprised between $Z=0.0004$ and $0.019$,
interpolating to generate stars of any intermediate 
metallicity.

The $IJK_{\mathrm s}$ photometry has been simulated from the
bolometric magnitudes by applying the extended tables of
bolometric corrections (BCs) from Girardi et al. (\cite{gi02}) for
O-rich stars (with C/O$<1$), and by empirical relations for C-rich
stars (with C/O$>1$). For O-rich stars, the magnitudes and colours 
correspond to the DENIS $IJK_{\mathrm s}$ and 2MASS
$JHK_{\mathrm s}$ filters; they are based on ATLAS9 model atmospheres 
(Castelli et al. \cite{cast}) for $\Teff\ga4000$~K, and on the
empirical spectra and \Teff--colour scale by Fluks et al.
(\cite{fluk}) for cooler stars. 
For C-type stars, the BC in the $K$ band is taken from the
relation by Costa \& Frogel (\cite{cost}), whereas the $(J-K)$ colour 
is derived from the \Teff--(\jk)--C/O relation from Marigo et al.
(\cite{ma03}), which itself is based on a fit to the empirical
data by Bergeat et al. (\cite{berg}).
Photometric errors were simulated as in Paper II.

More details about the properties of
theoretical models as well as the number and distribution
of AGB stars for varying metallicity at a constant SFR, and
viceversa, are given in Paper II. For the sake of simplicity, 
we adopt a simple family of exponentially increasing/decreasing
SFRs, $\psi(t)\propto\exp(t/\alpha)$, where $t$ is the stellar age,
and $\alpha$ a free parameter. Table~1 of Paper~II presents 
the correspondence between $\alpha$ and the mean age of all stars 
formed in a model.
Figure \ref{fig_CM_SFR} shows the variation of the C/M ratio as a
function of [M/H] for different cases of SFR. This ratio has been
calculated in two different ways: first using the numbers found from a
given model, and second from the number of stars falling in the C-rich 
 and O-rich AGB regions in the near-IR CMD (see Paper II). 
These two selection criteria (the ``model'' and ``photometric'' ones, 
respectively) produce C/O ratios with a similar behaviour for
different [M/H] and SFR, as can be seen in Fig. \ref{fig_CM_SFR}.
The photometric criterion is the same as the observational one 
described in Sect.~\ref{obs}.

The distinction between AGB stars of a different chemical type is clear
in the CMD of Fig. \ref{fig_compMCs}. This figure shows a 
simulation that adopts the SFR and the AMR we use in the ``realistic''
case -- as derived from Harris \& Zaritsky (\cite{harr}) in the entire
body of the SMC, and from chemical evolution models 
(Pagel \& Tautvaisiene \cite{page}). The corresponding luminosity
functions (LFs) are presented in Fig.~\ref{fig_LFs}.

\section{Comparison between observed and theoretical distributions}
\label{comp}

Although the photometric criteria for selecting C and M AGB stars
provide LFs in good agreement with the model distribution
(Fig. \ref{fig_LFs}) the number of M-type AGB stars predicted by the
models is just about half of the observed ones. This was also the case
in the LMC and it is either due to inappropriate SFR and AMR 
for the investigated AGB population or to uncorrected lifetimes of the C- and 
O-rich phases of TP-AGB evolution. Foreground contamination is negligible.
Similarly to Paper II we proceed by fixing the 
AMR relation of the SMC, and change the family of SFR. 
Then for each simulation we compare, using the $\chi^2$ test, the derived and 
observed LF, separately for C and M stars in each sector of each ring. 
Figure \ref{fig_LFs} shows a model (left panel) that represents 
rather well the distribution of C stars but fails to recover the distribution 
of M stars while on the right panel both distributions are recovered with a 
probability above $85$\%. For the sake of simplicity, we adopt a constant AMR, 
i.e. the same $Z$ for all ages. This is not the best approximation
for the case of the SMC, since both cluster and field data
indicate that a substantial chemical enrichment occurred in the last 
3~Gyr of its history (Harris \& Zaritsky 2004, and references therein). 
However, the adoption of a single reference $Z$ is necessary to
unequivocably locate the lines that are used to 
select C and M stars in the CMD.

\begin{figure*}
\hspace{-0.6cm}
\epsfxsize=0.24\hsize \epsfbox{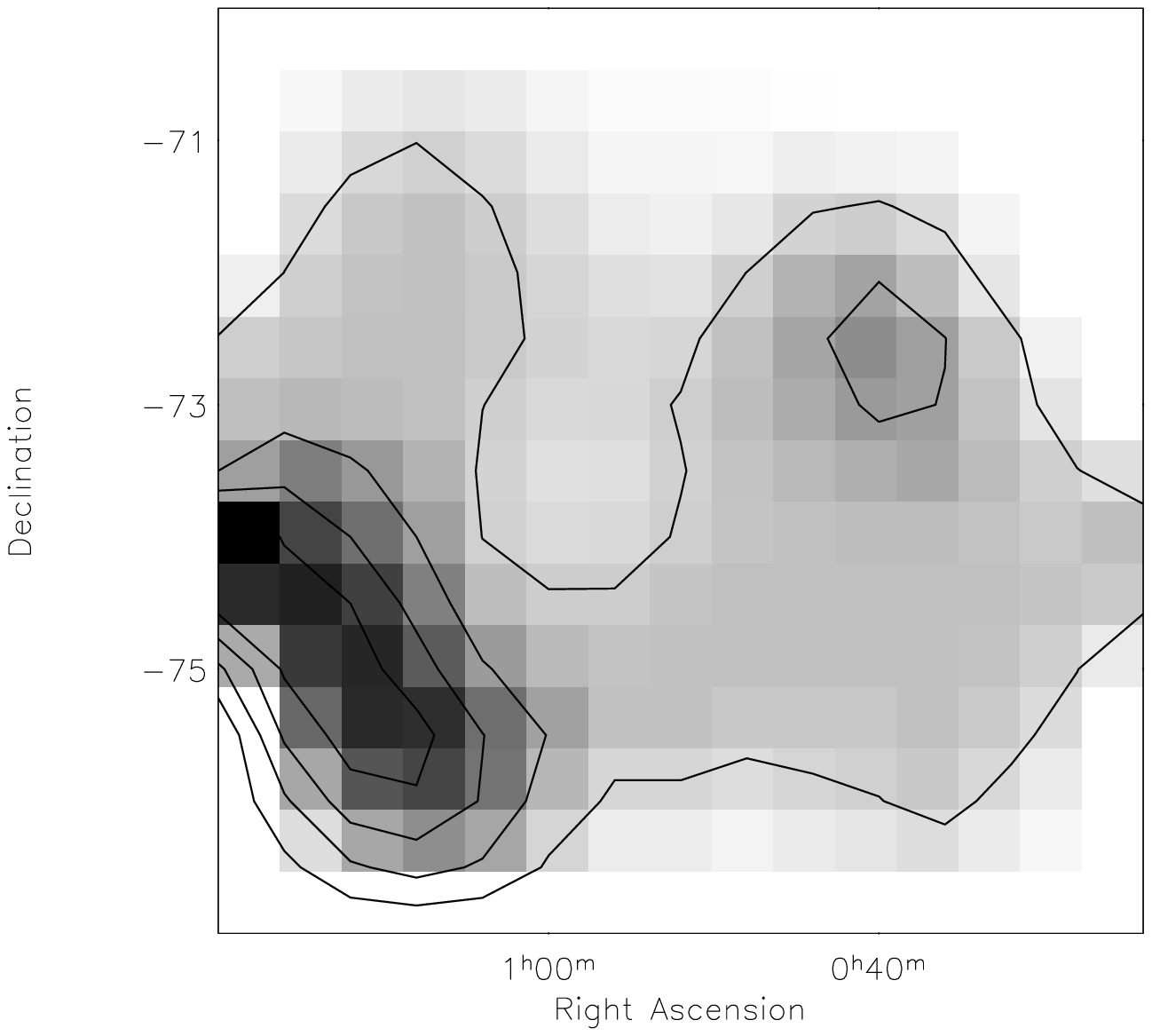}
\hspace{-1.0cm}
\epsfxsize=0.24\hsize \epsfbox{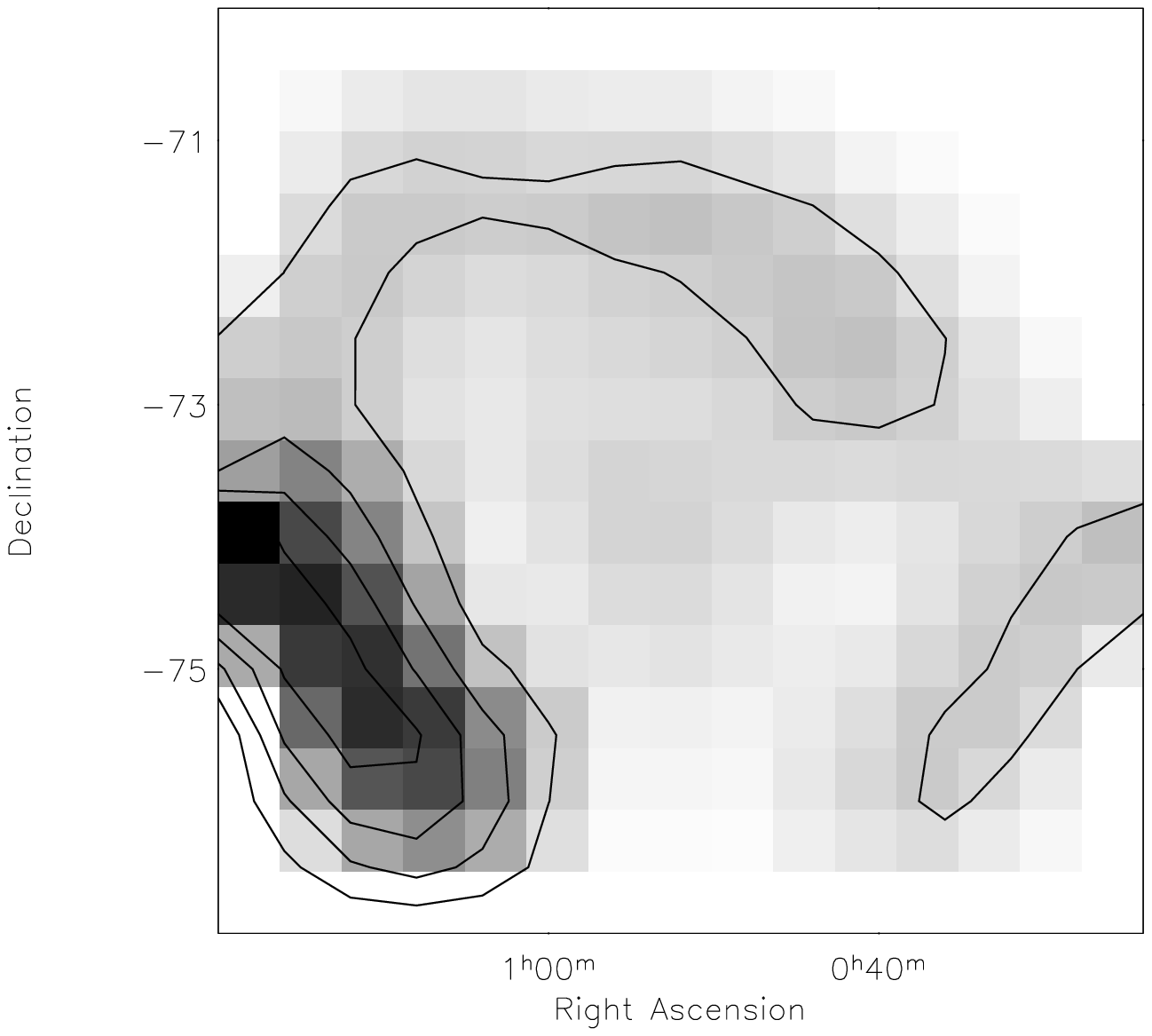}
\hspace{-1.0cm}
\epsfxsize=0.24\hsize \epsfbox{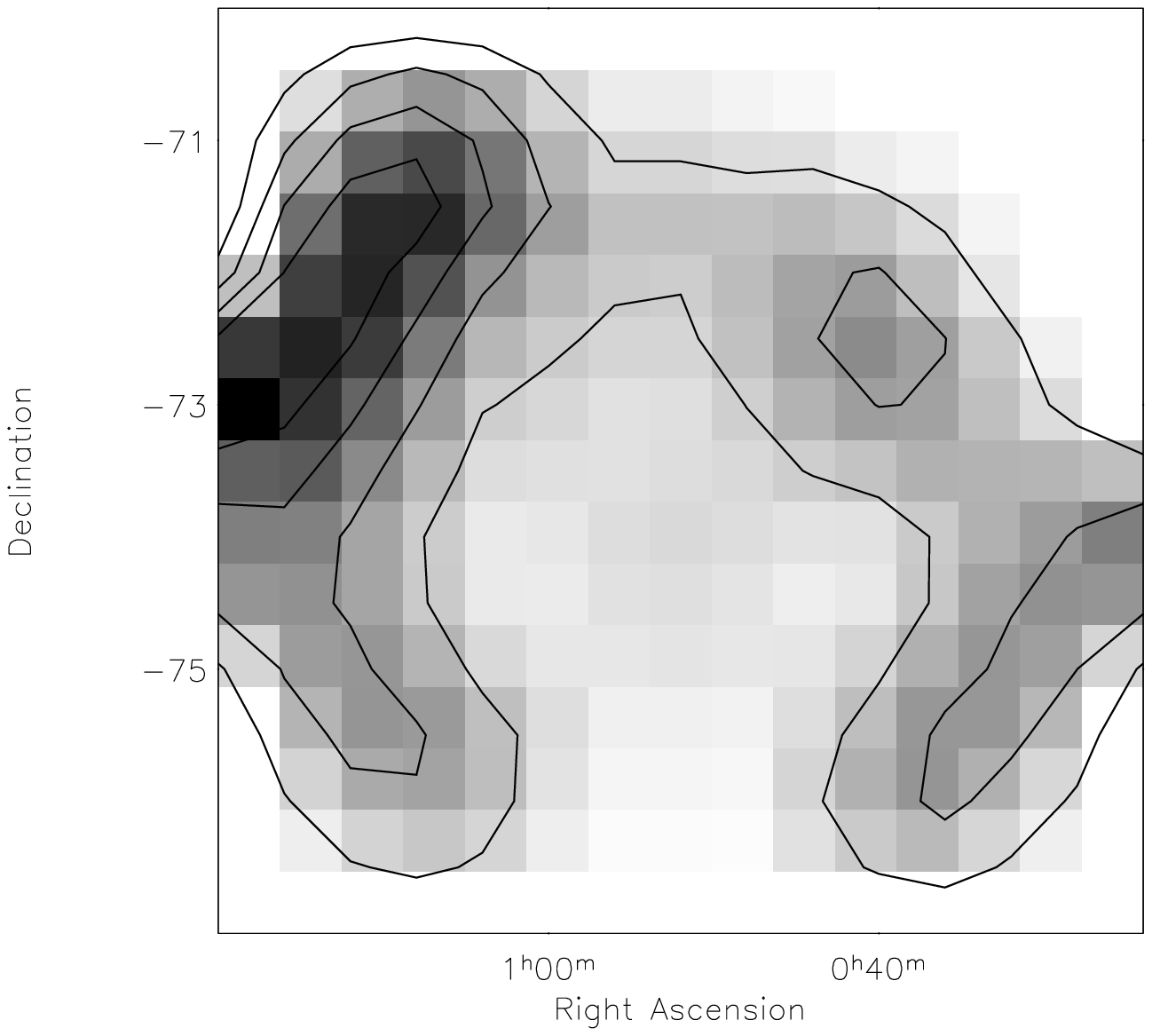}
\hspace{-1.0cm}
\epsfxsize=0.24\hsize \epsfbox{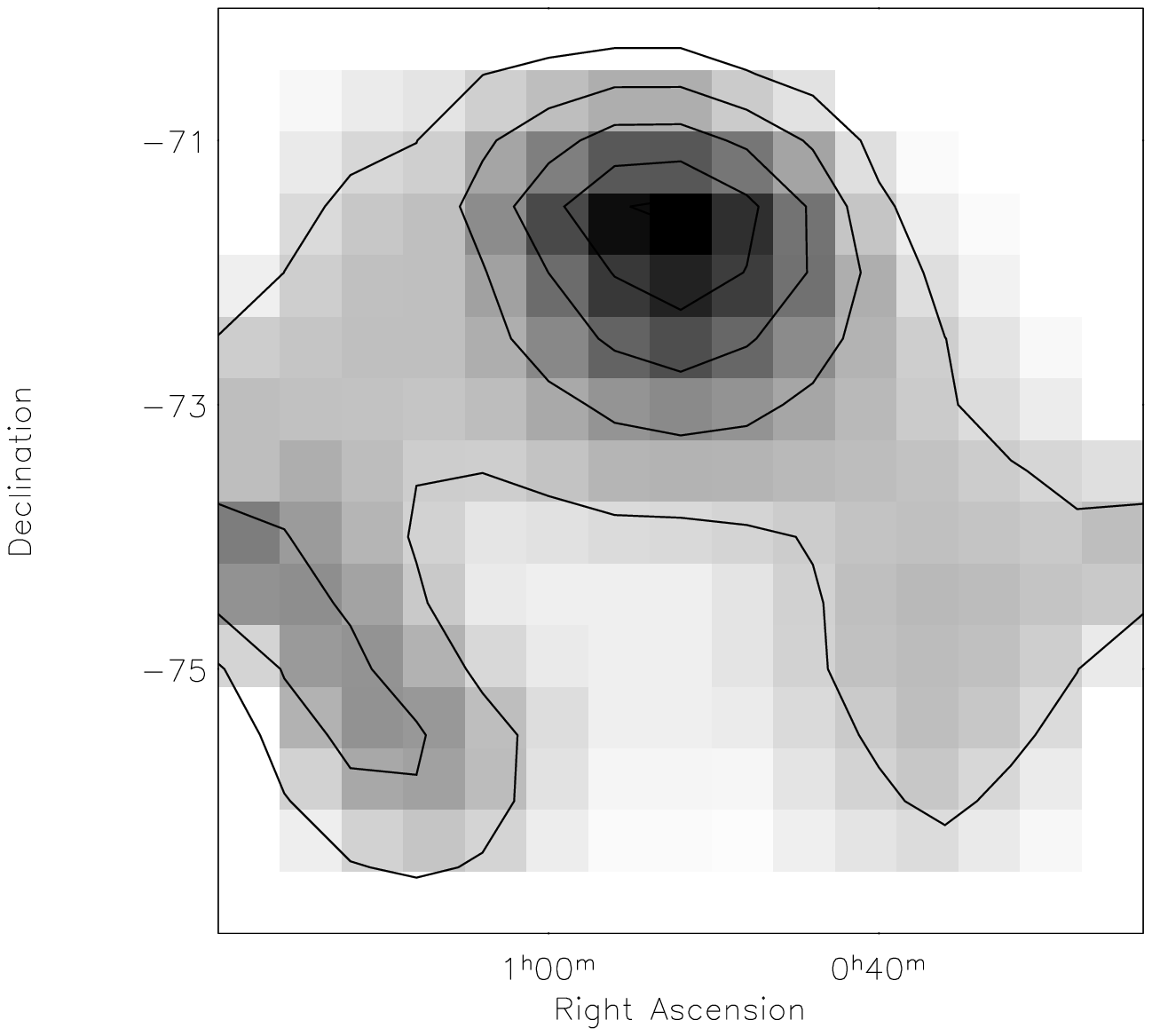}
\hspace{-1.0cm}
\epsfxsize=0.24\hsize \epsfbox{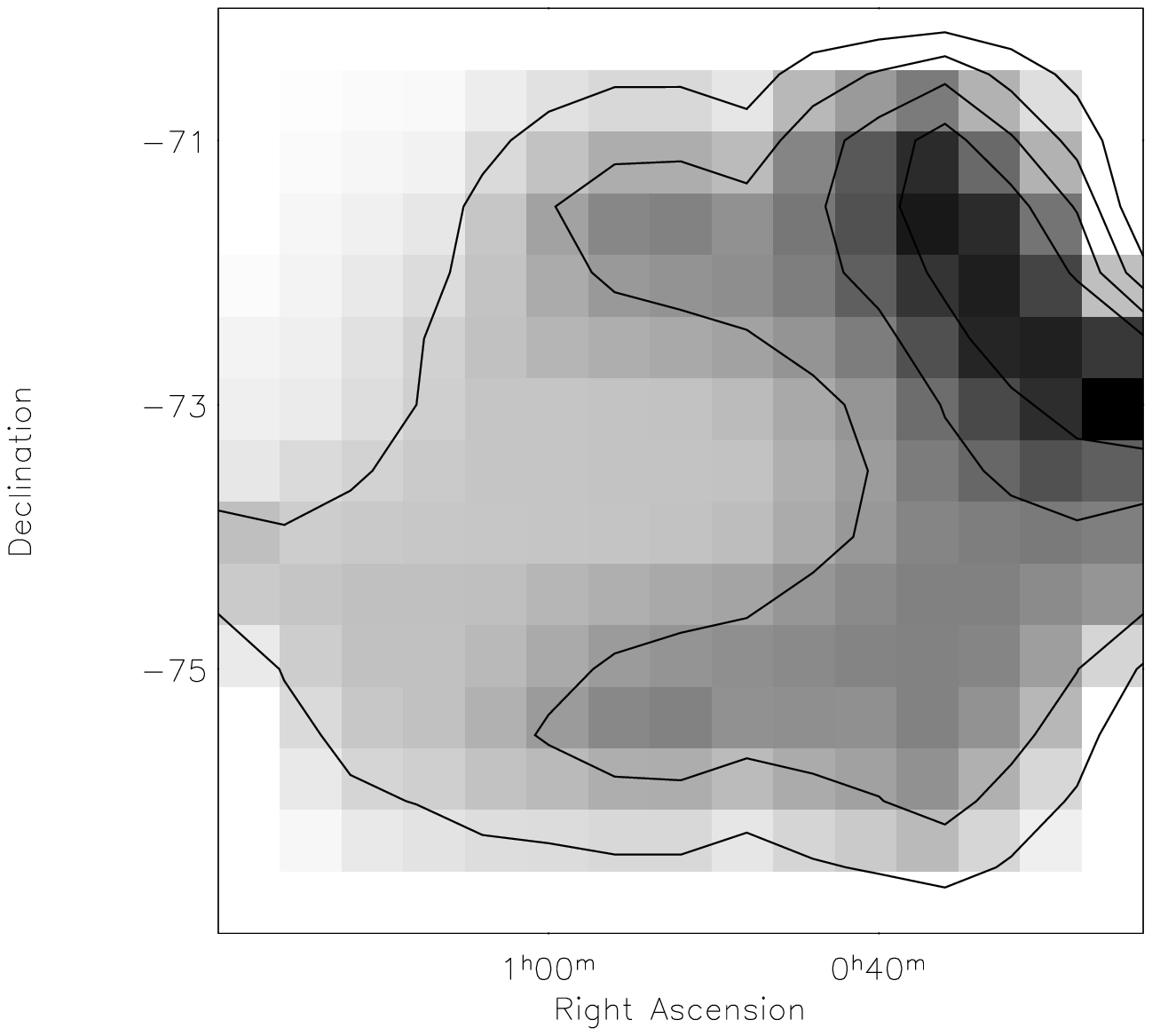}

\hspace{-0.6cm}
\vspace{-0.1cm}
\epsfxsize=0.24\hsize \epsfbox{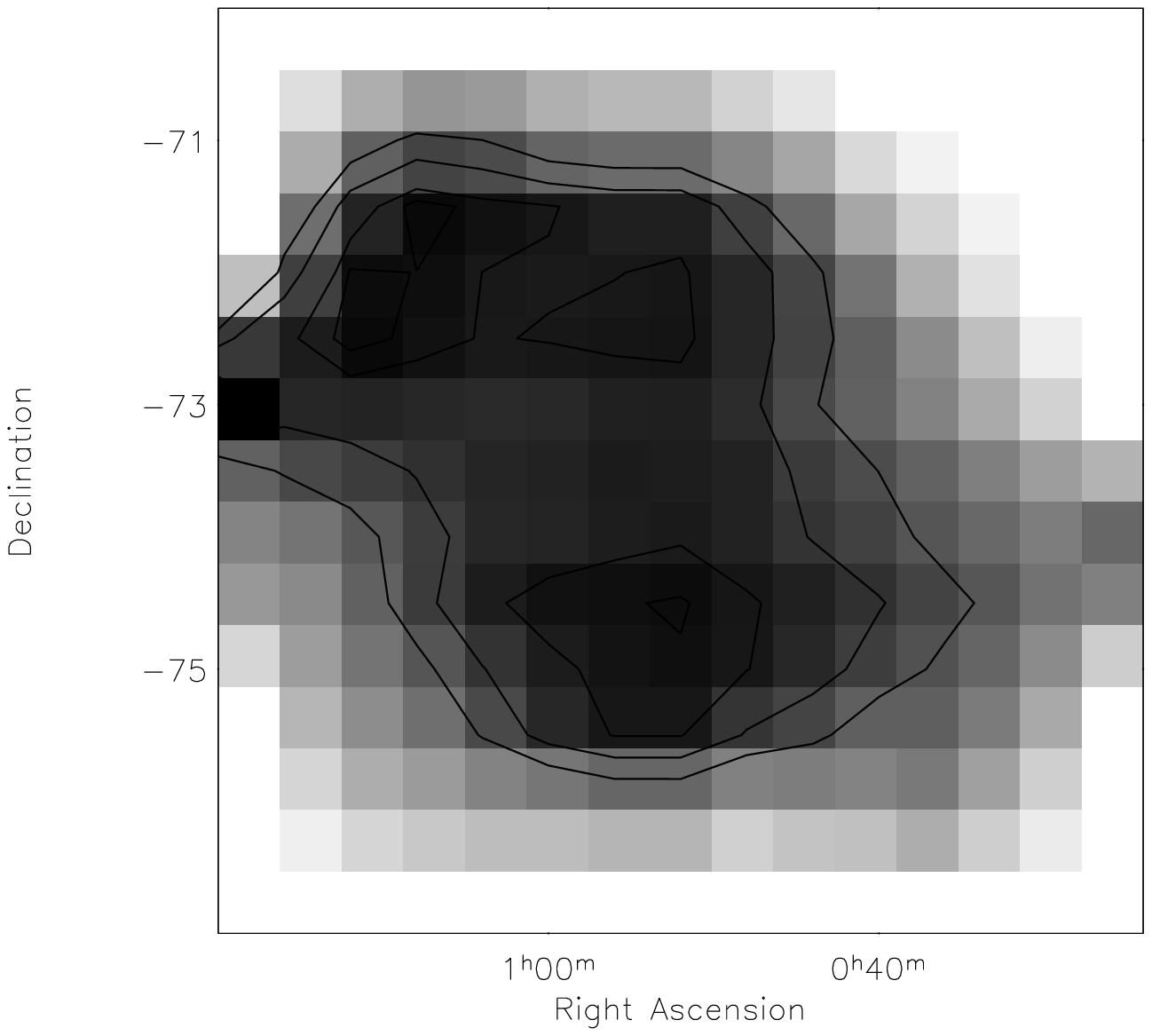}
\hspace{-1.0cm}
\epsfxsize=0.24\hsize \epsfbox{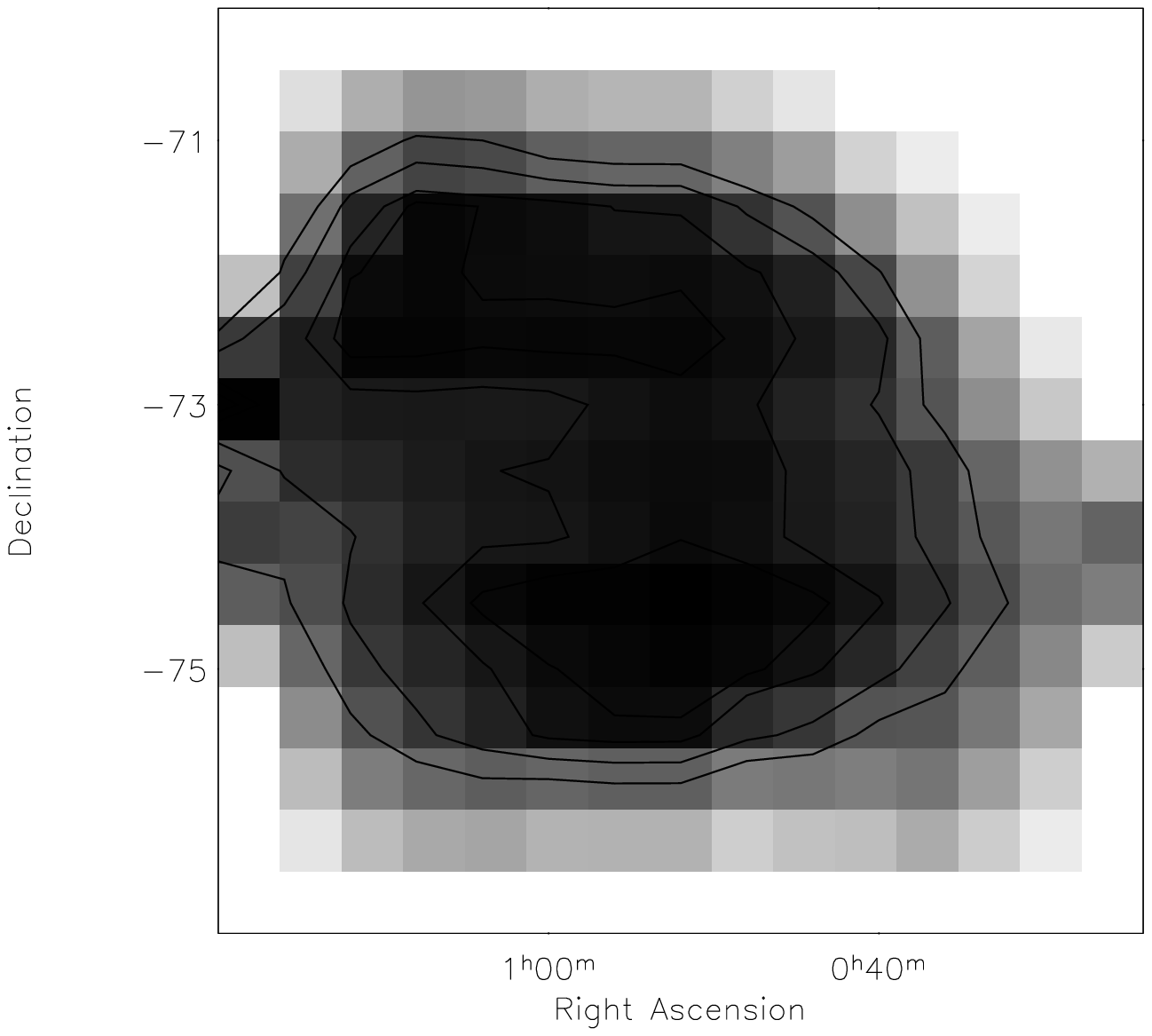}
\hspace{-1.0cm}
\epsfxsize=0.24\hsize \epsfbox{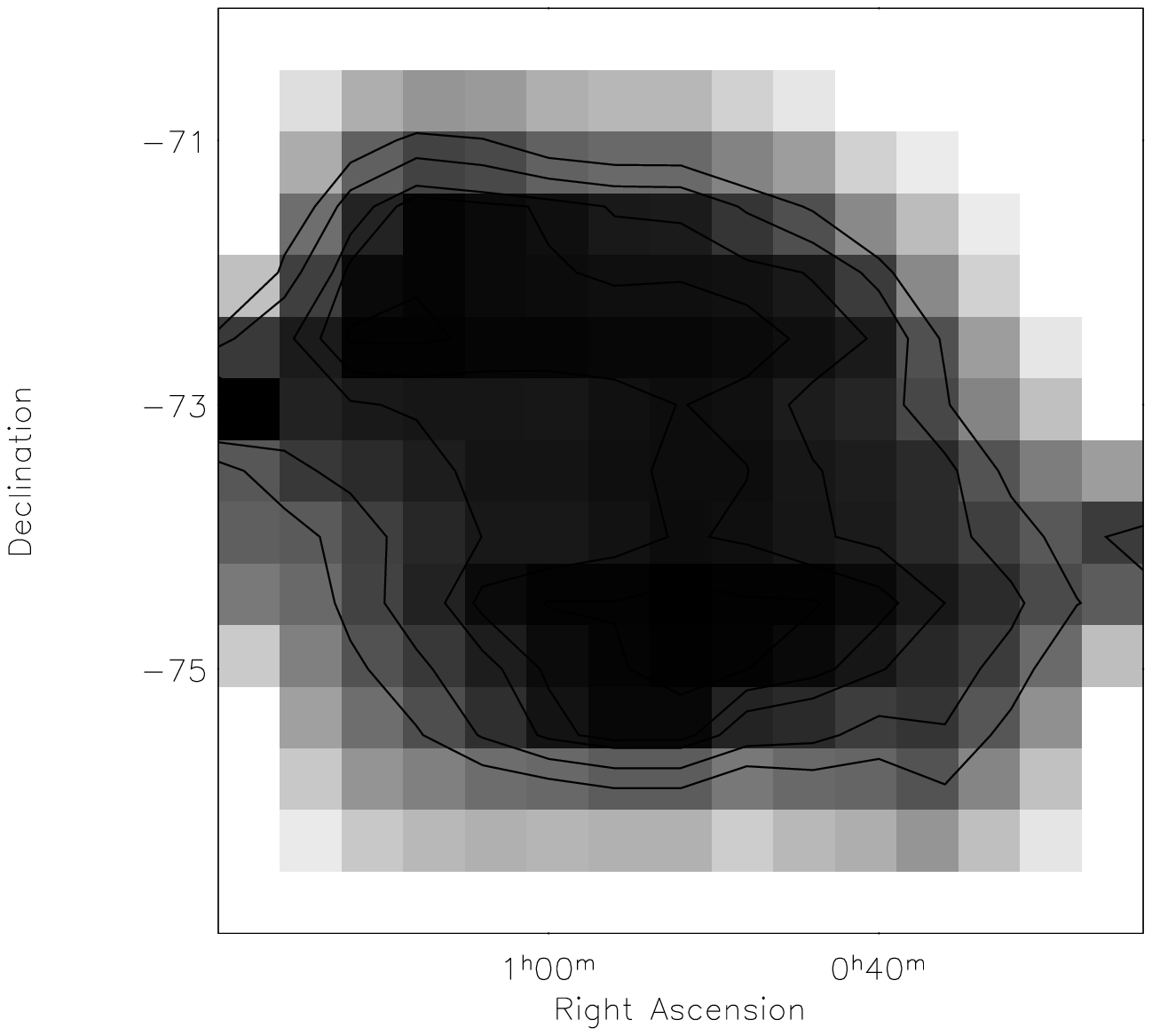}
\hspace{-1.0cm}
\epsfxsize=0.24\hsize \epsfbox{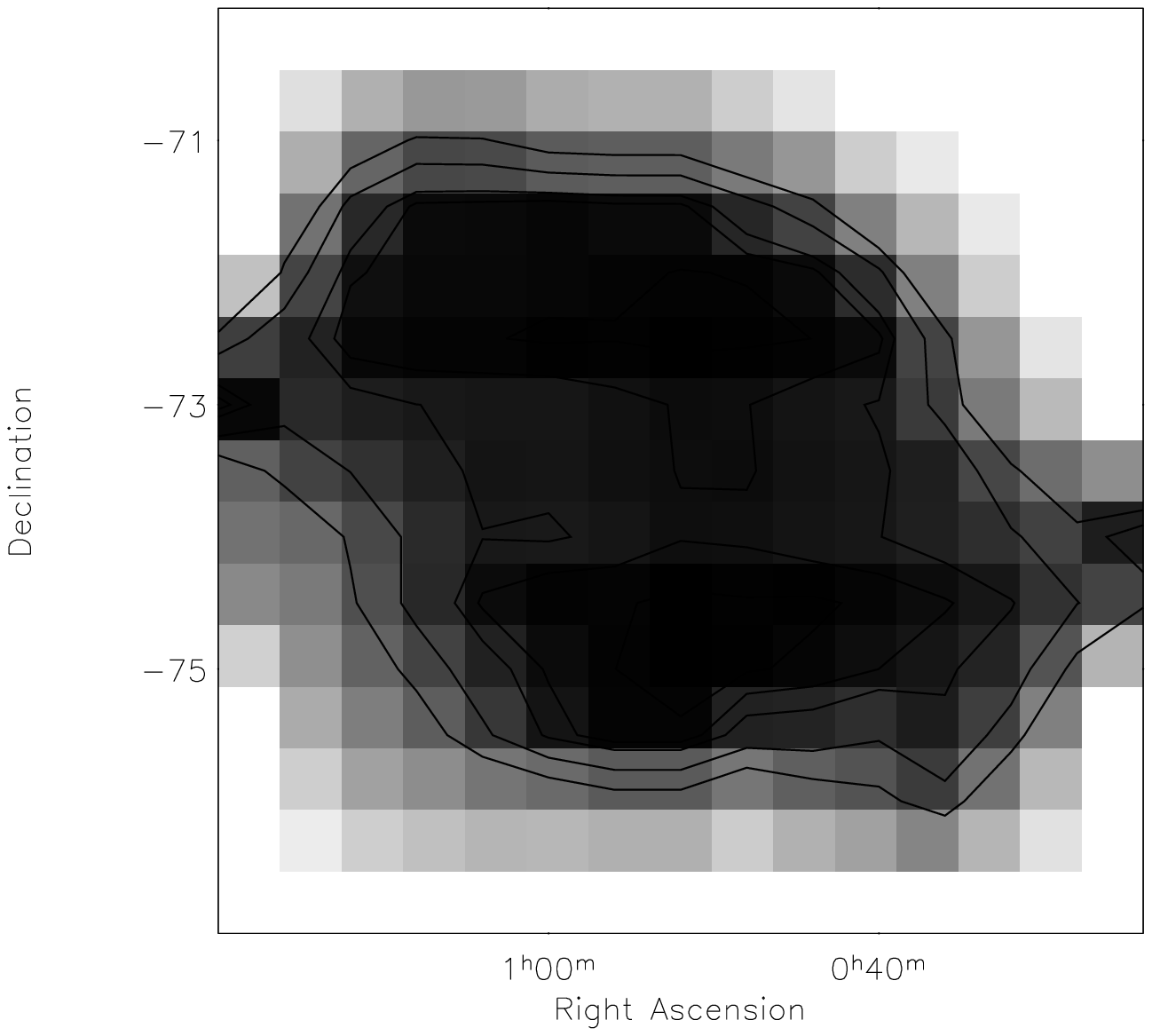}
\hspace{-1.0cm}
\epsfxsize=0.24\hsize \epsfbox{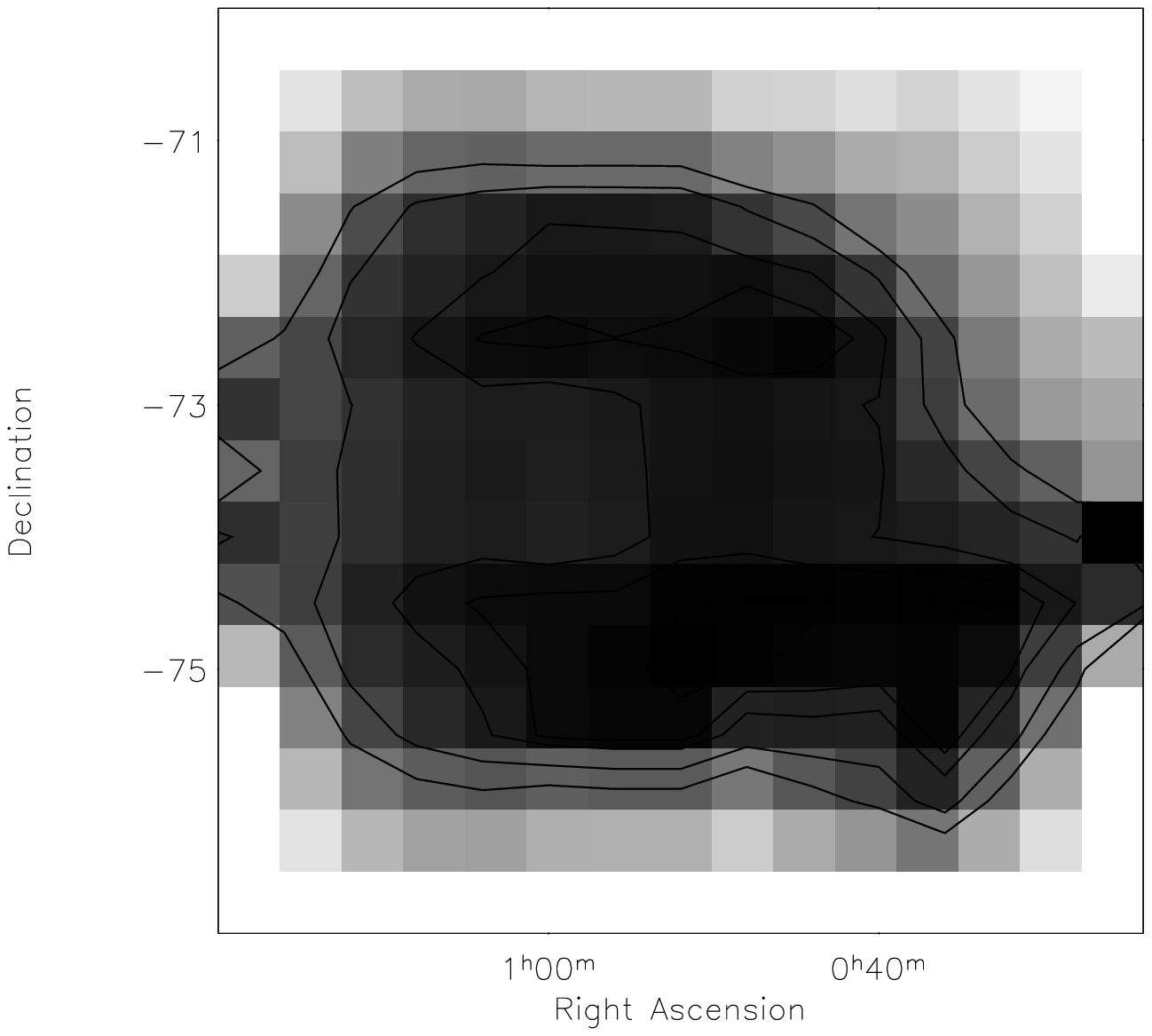}

\hspace{-0.6cm}
\vspace{-0.1cm}
\epsfxsize=0.24\hsize \epsfbox{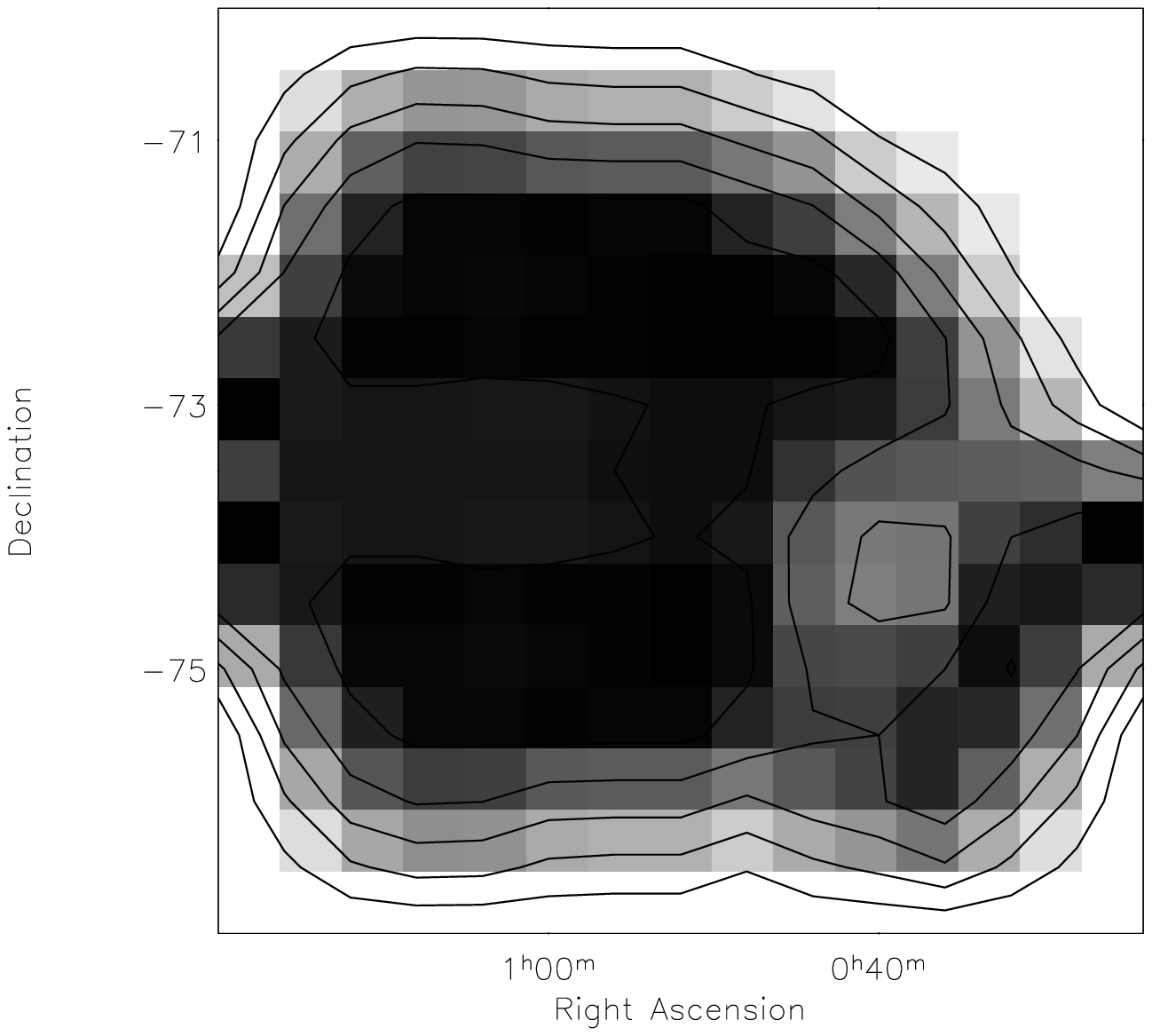}
\hspace{-1.0cm}
\epsfxsize=0.24\hsize \epsfbox{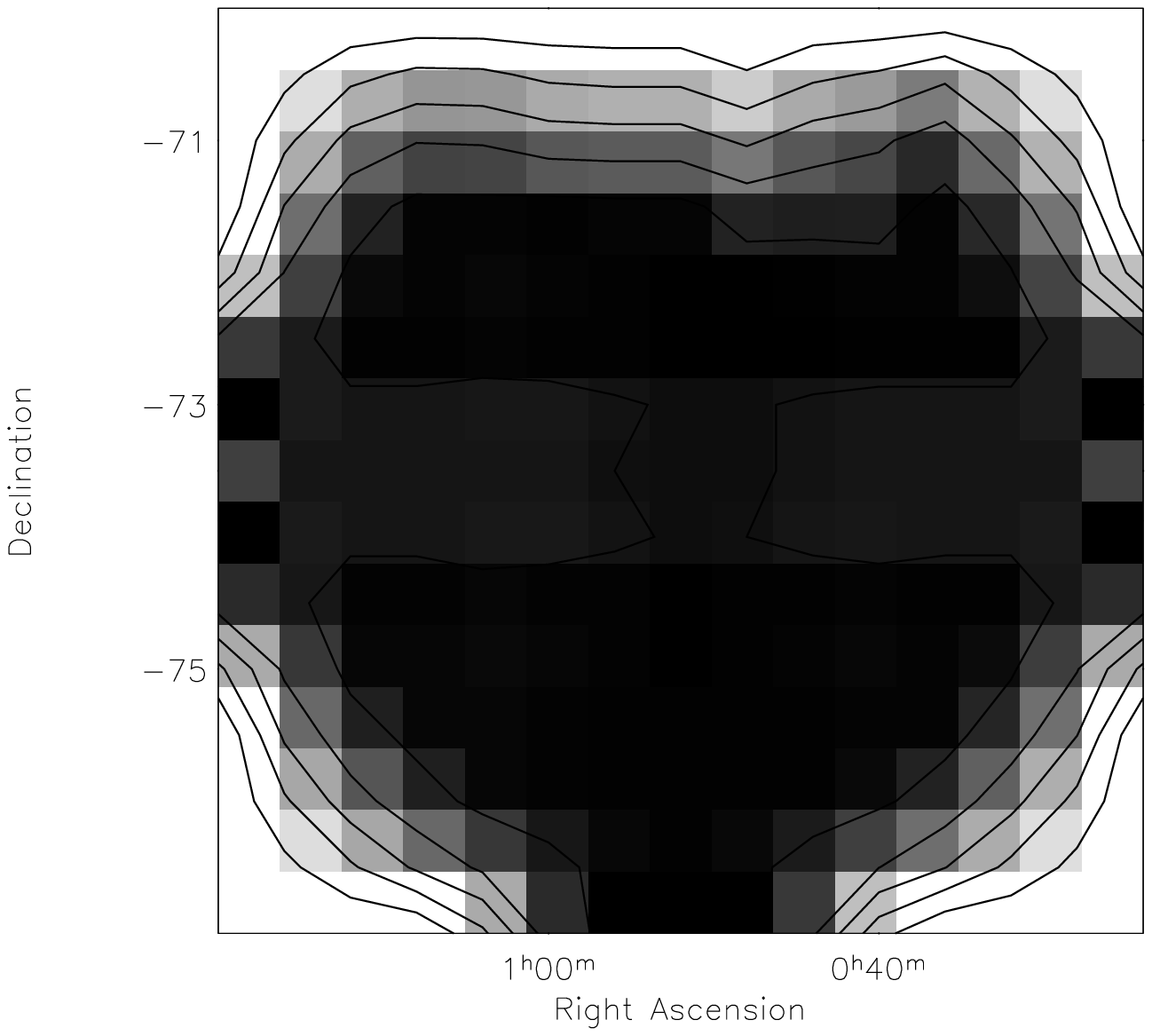}
\hspace{-1.0cm}
\epsfxsize=0.24\hsize \epsfbox{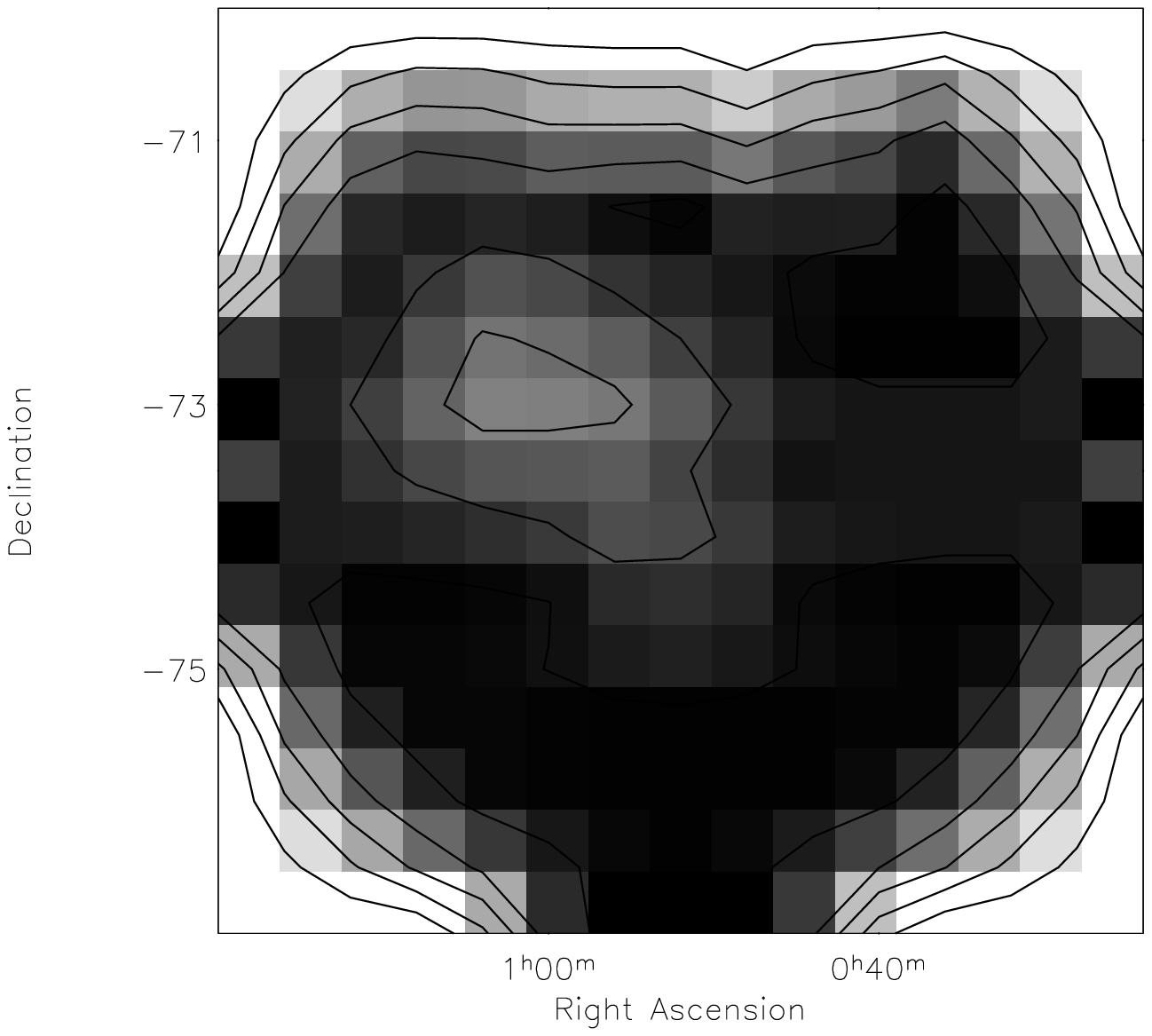}
\hspace{-1.0cm}
\epsfxsize=0.24\hsize \epsfbox{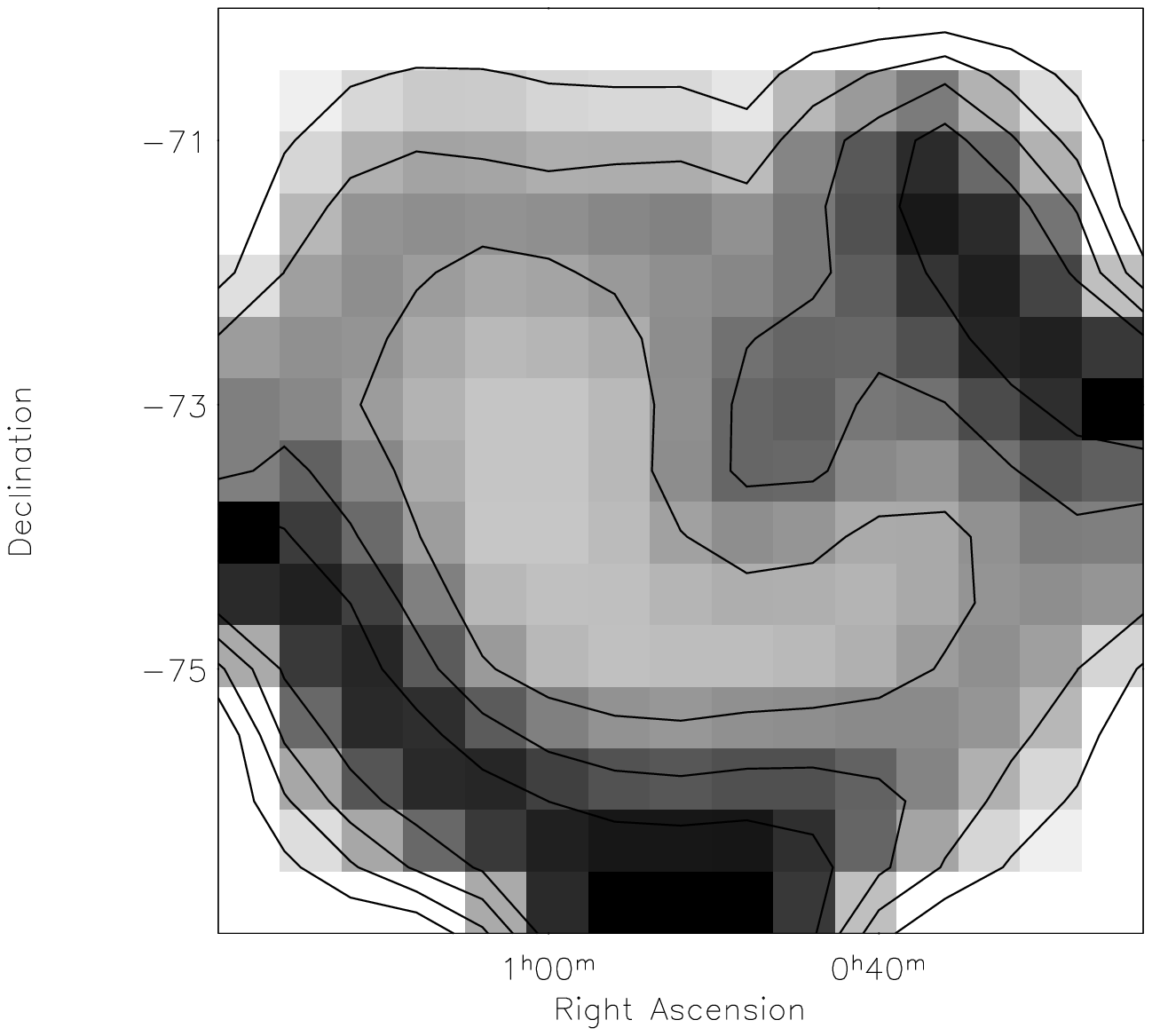}
\hspace{-1.0cm}
\epsfxsize=0.24\hsize \epsfbox{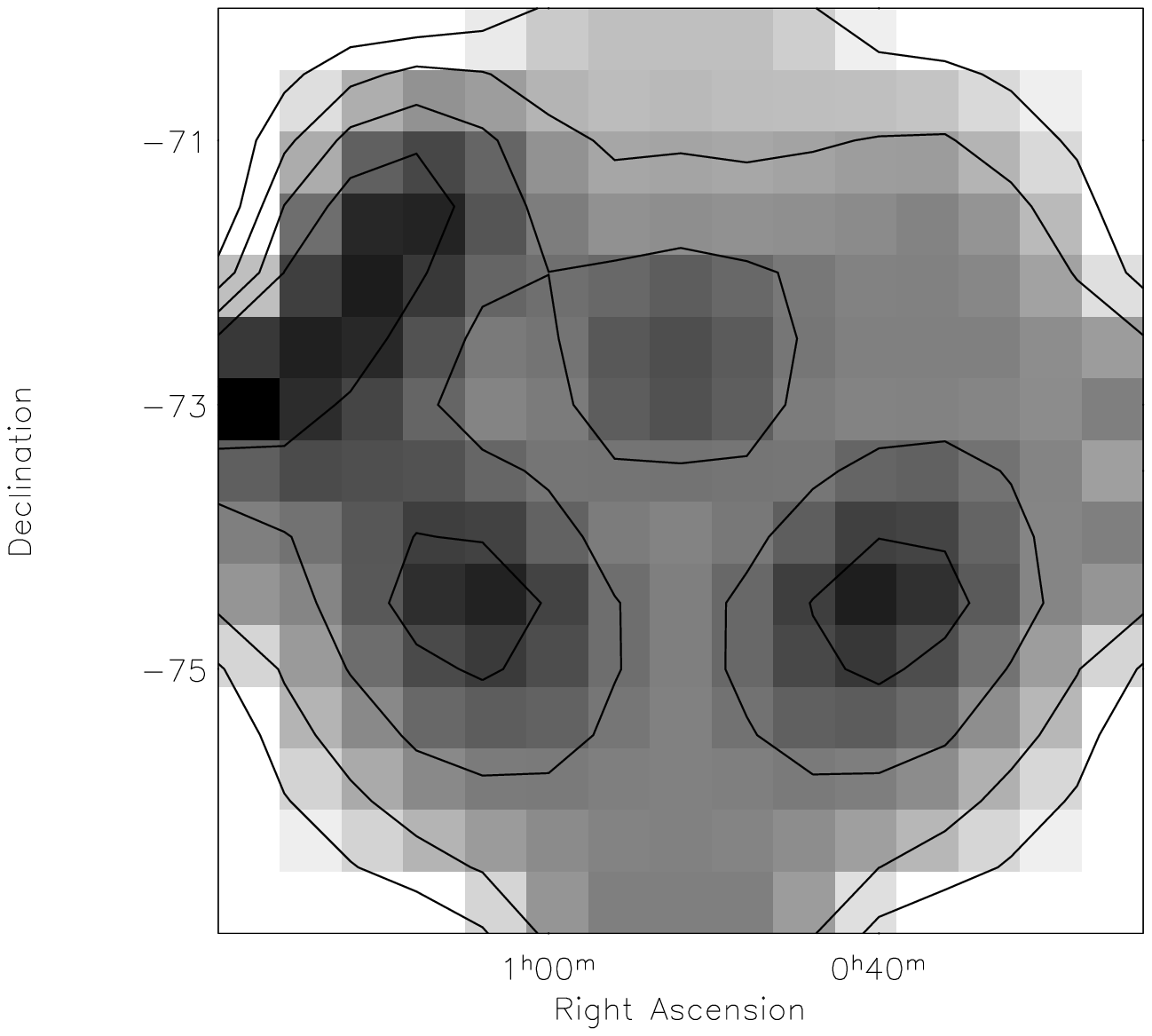}

\hspace{-0.6cm}
\vspace{-0.1cm}
\epsfxsize=0.24\hsize \epsfbox{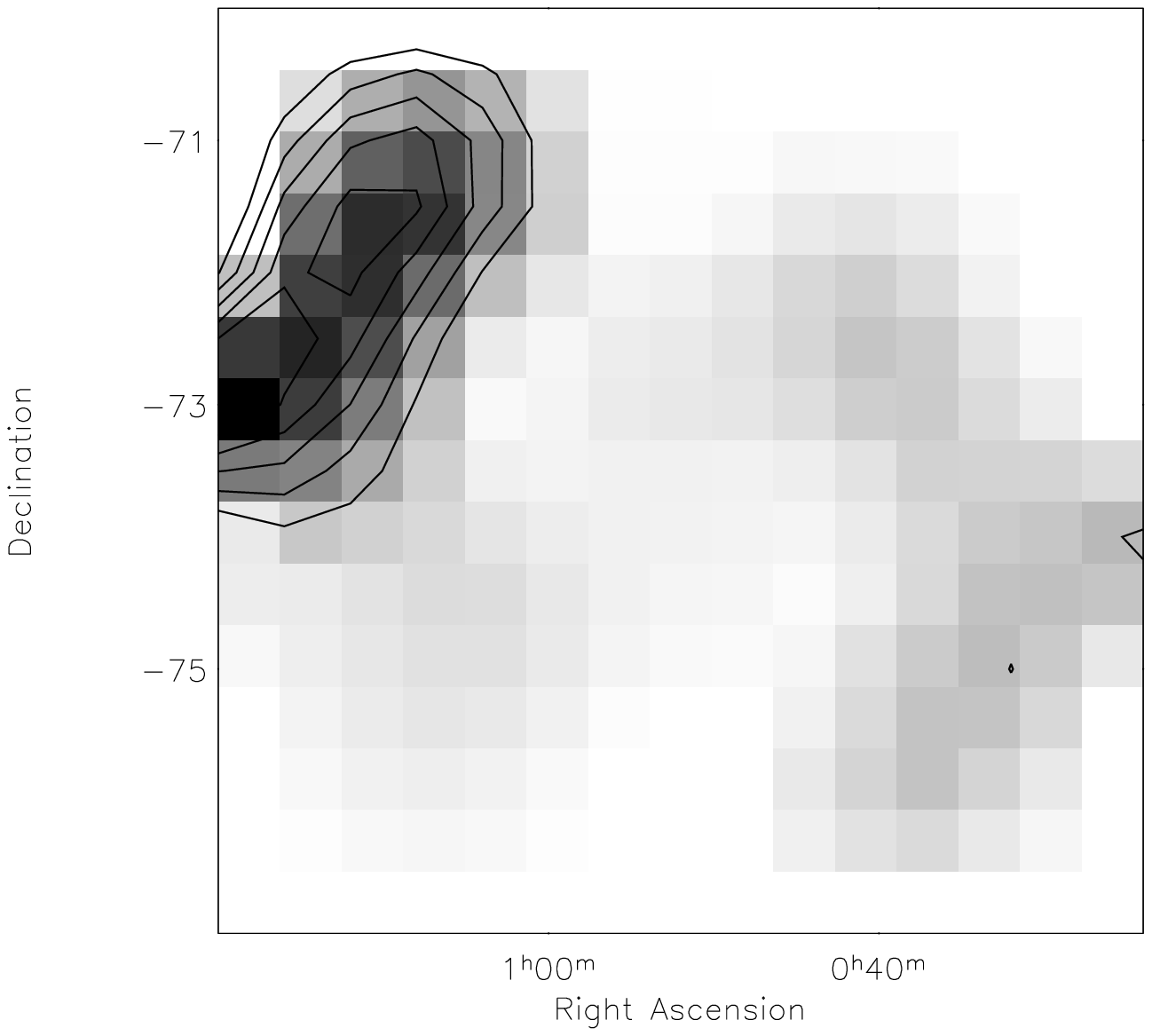}
\hspace{-1.0cm}
\epsfxsize=0.24\hsize \epsfbox{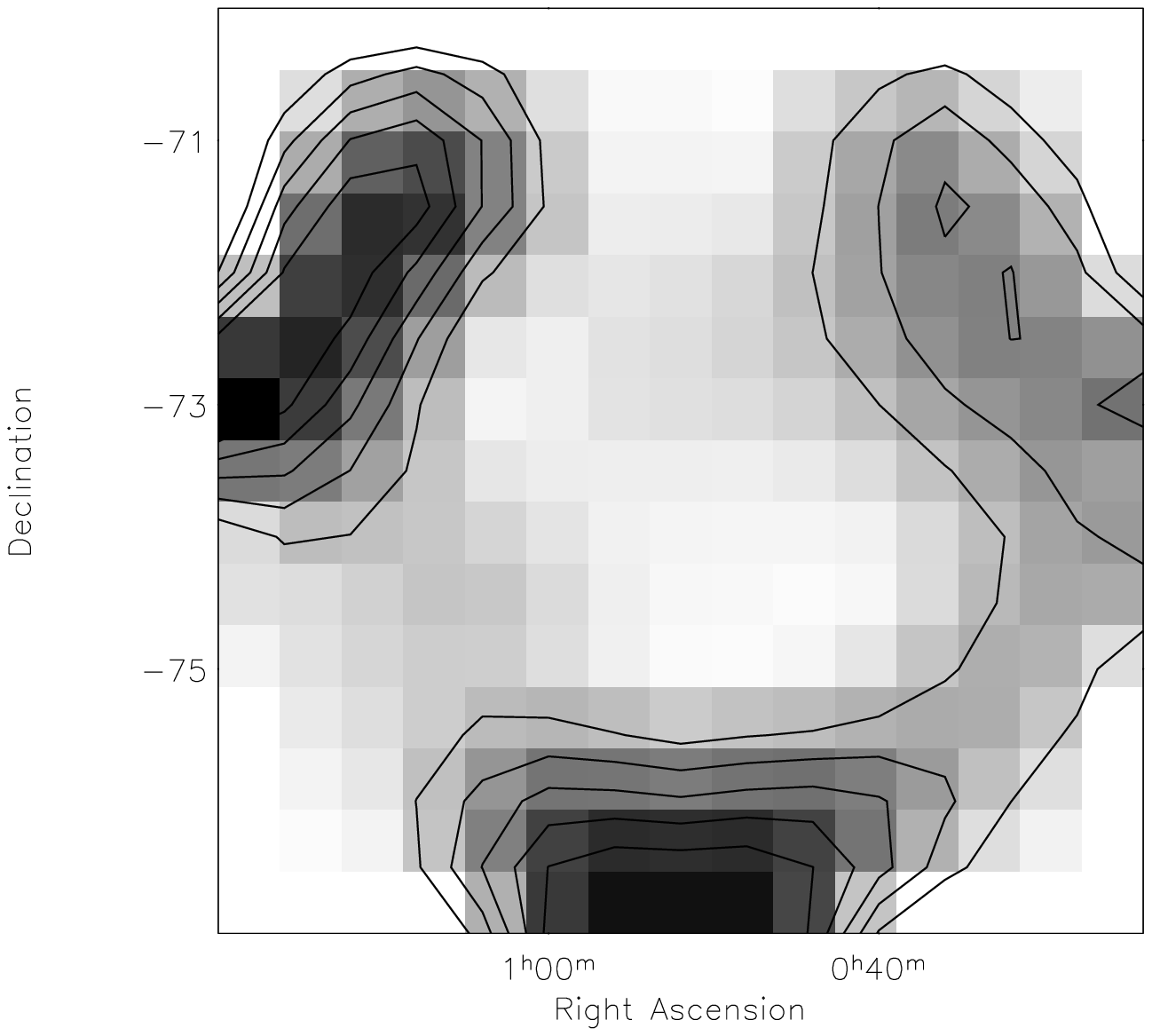}
\hspace{-1.0cm}
\epsfxsize=0.24\hsize \epsfbox{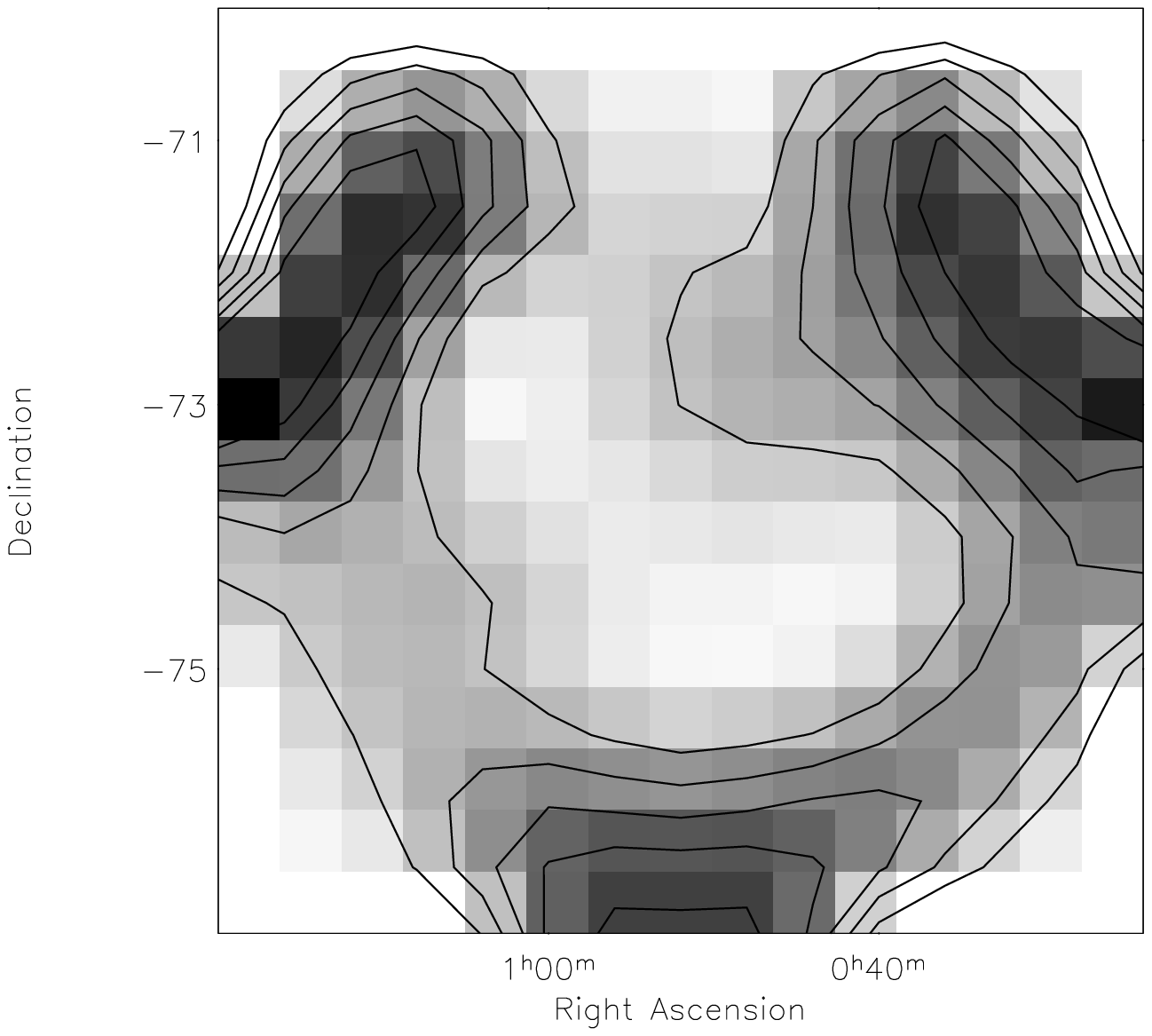}
\hspace{-1.0cm}
\epsfxsize=0.24\hsize \epsfbox{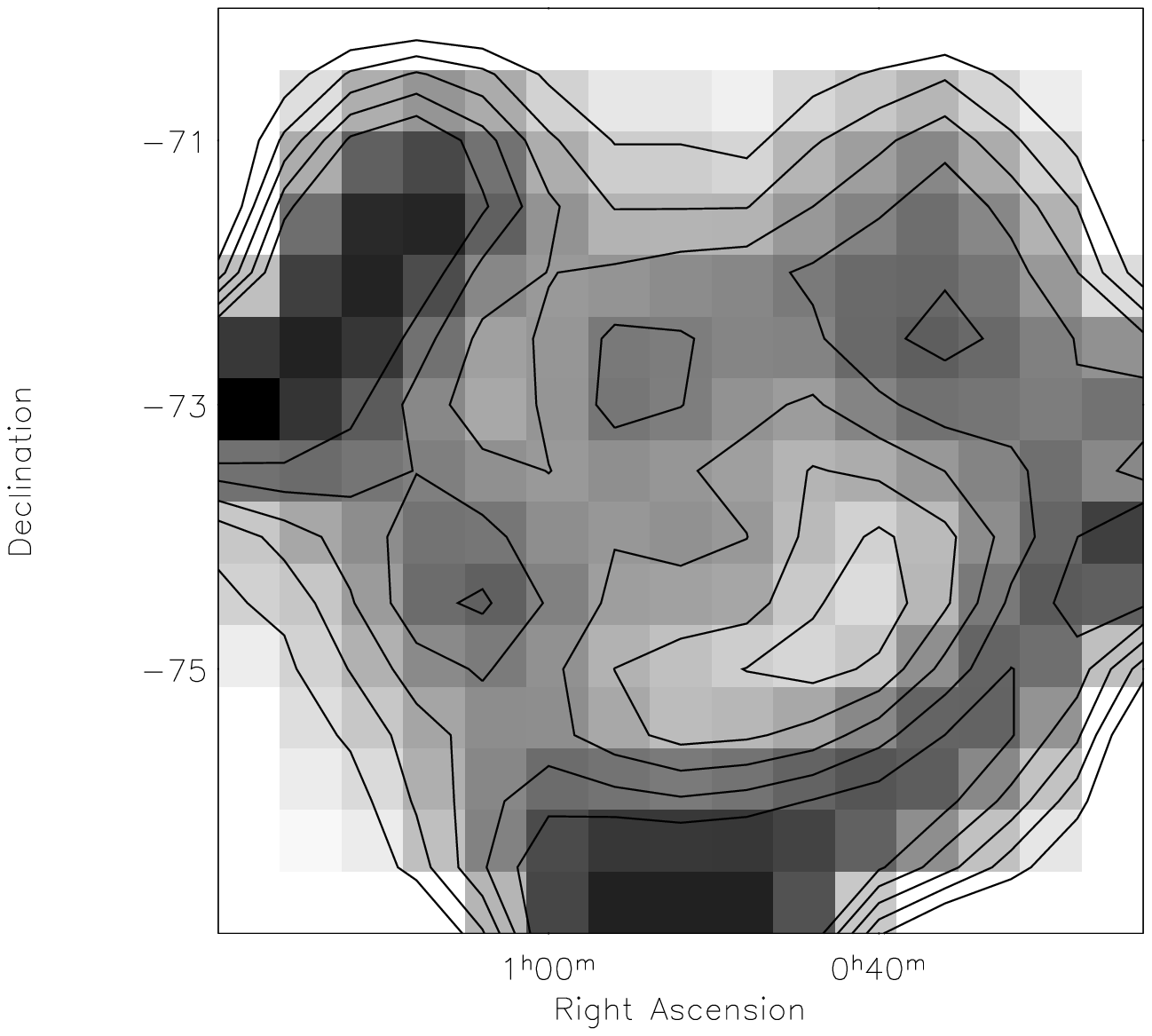}
\hspace{-1.0cm}
\epsfxsize=0.24\hsize \epsfbox{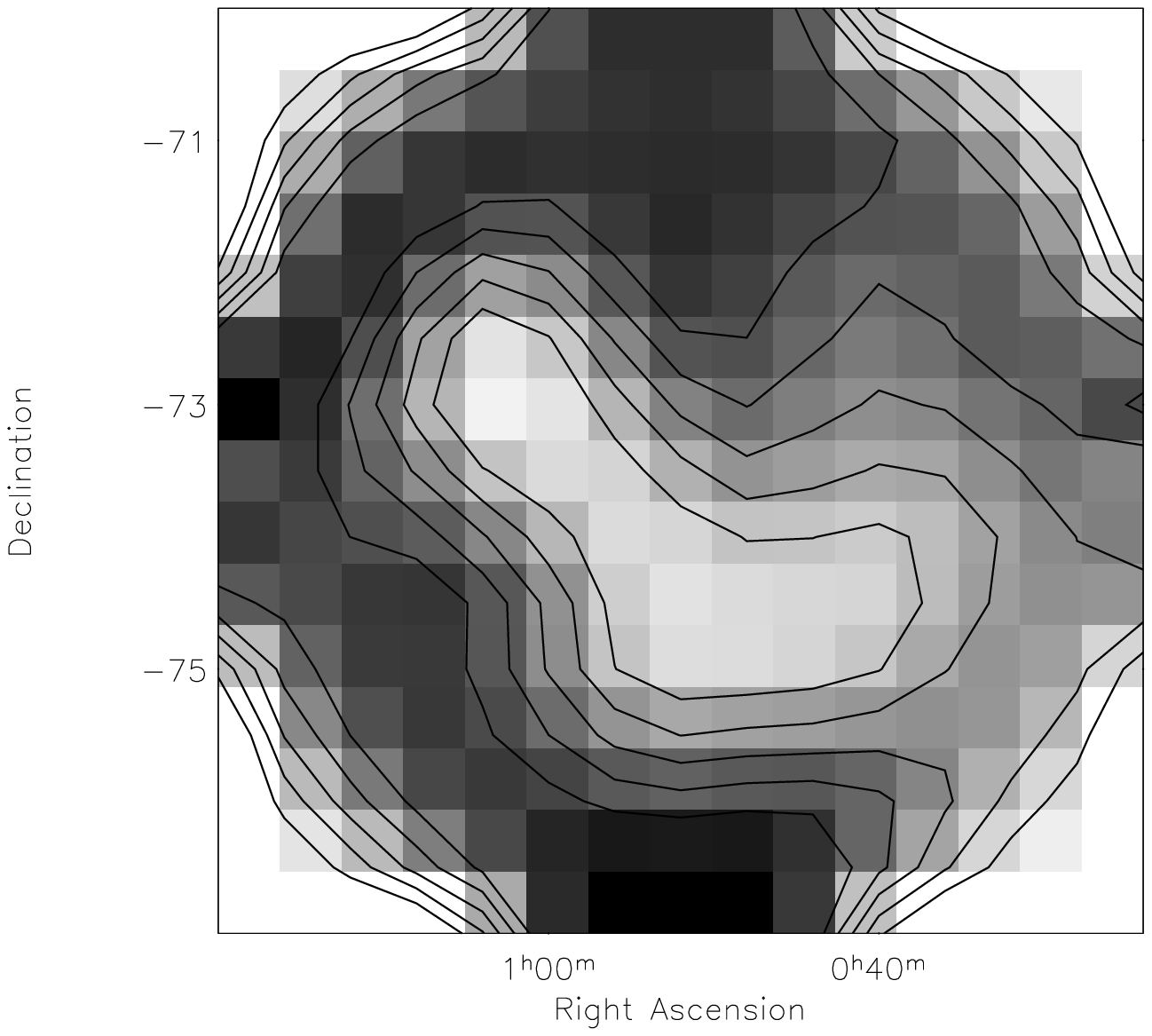}
\caption{Distribution of the metallicity that corresponds to the
maximum probability for a given SFR for C stars ({\bf first row}) and M stars
({\bf third row}). From left to right the SFR corresponds to a
population with a mean age of: $2$, $3.9$, $6.3$, $8.7$ and $10.6$
Gyr. Metallicity contours are at $0.003-0.015$ with a step $0.003$. 
Darker regions correspond to higher values. The
probability distributions corresponding to each SFR are shown in the 
{\bf second row} for C stars (contours are at $0.7$, $0.8$, $0.9$,
$0.94$ and $0.98$) and in the {\bf fourth row} for M stars (contours
are at $0.2$, $0.3$, $0.4$, $0.5$ and $0.6$).} 
\label{lumall}
\end{figure*}

\begin{figure*}
\begin{minipage}{0.65\hsize}
\hspace{-0.6cm}
\epsfxsize=0.39\hsize \epsfbox{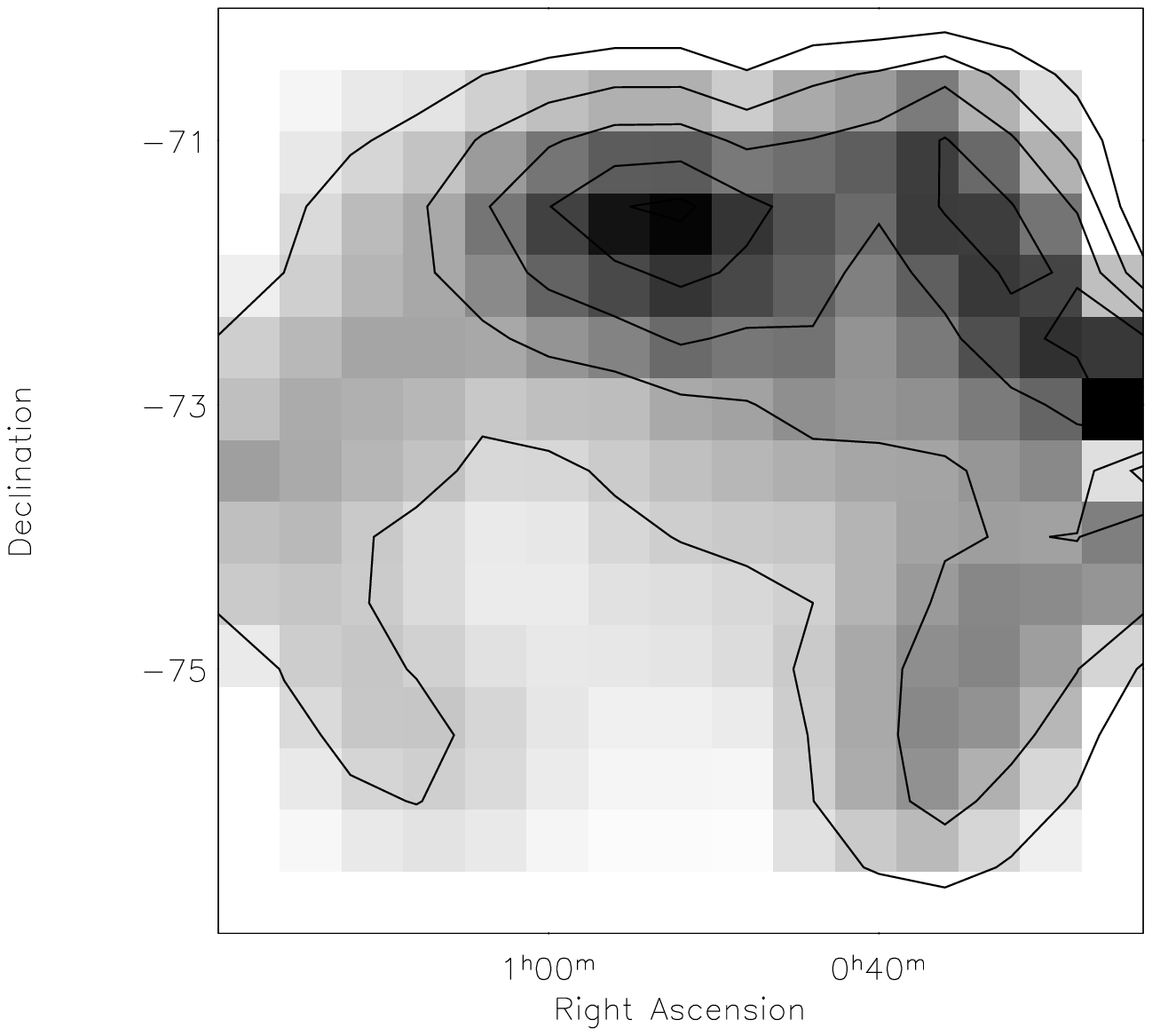}
\hspace{-1.4cm}
\epsfxsize=0.39\hsize \epsfbox{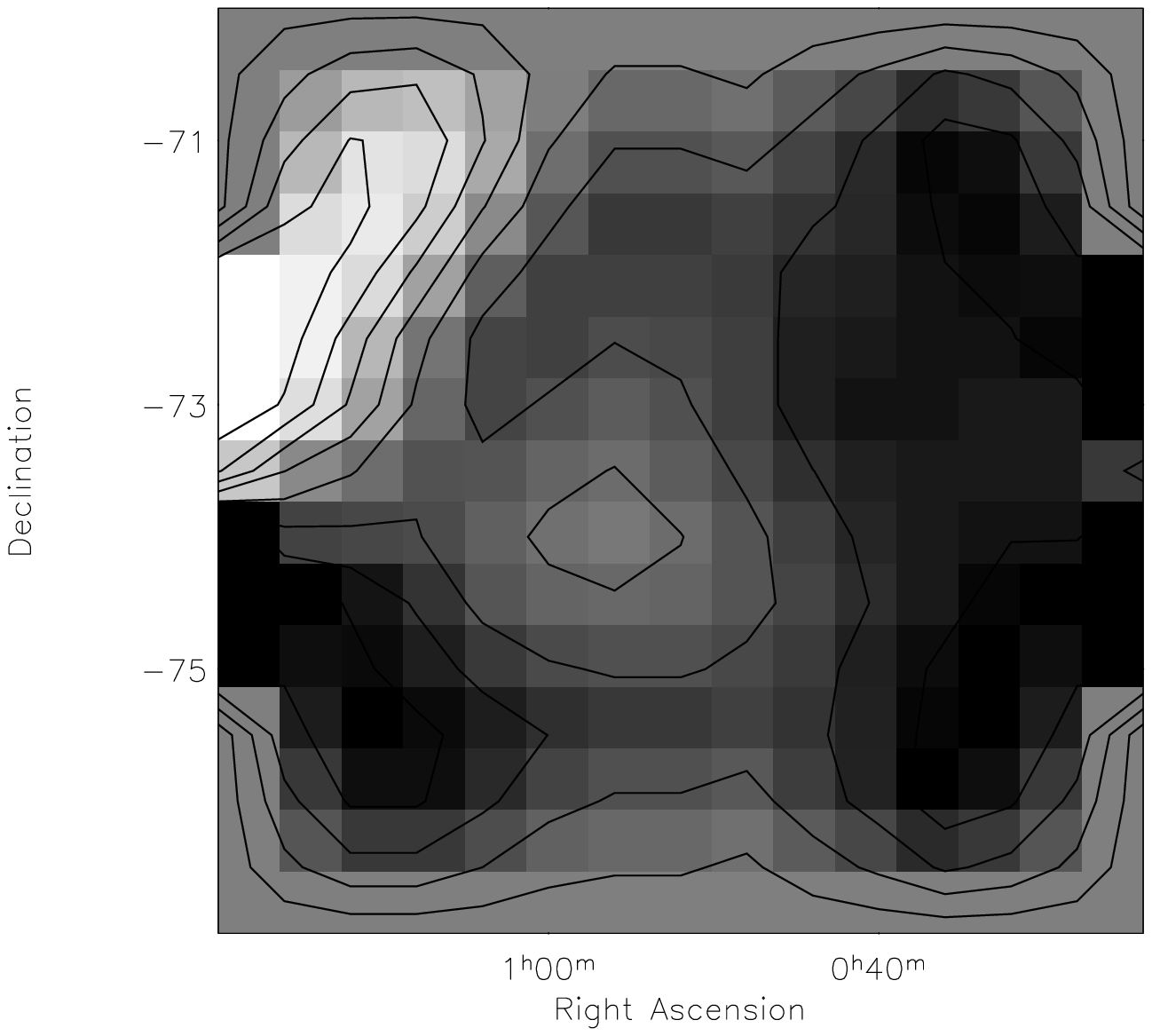}
\hspace{-1.4cm}
\epsfxsize=0.39\hsize \epsfbox{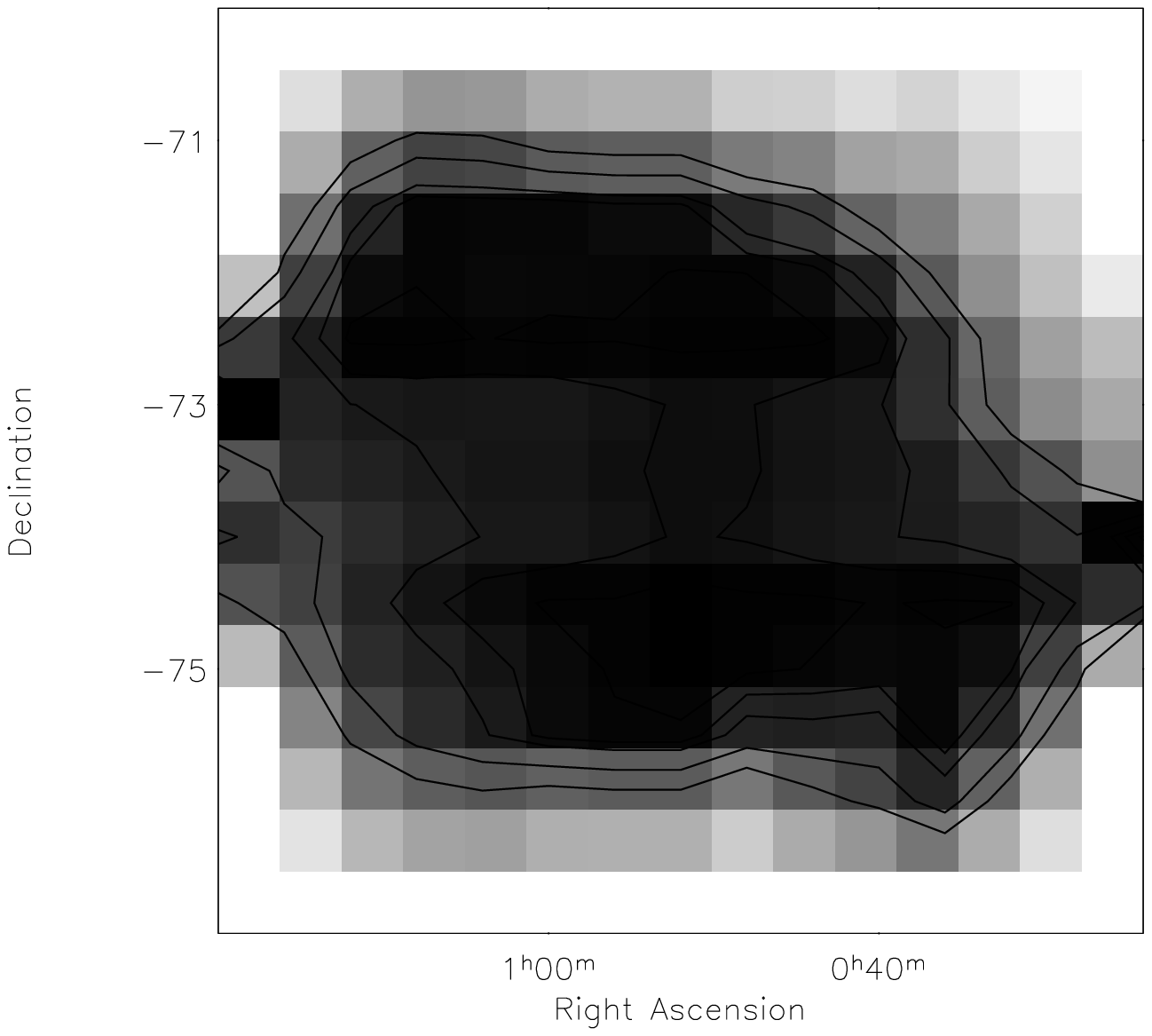}

\hspace{-0.6cm}
\vspace{-0.1cm}
\epsfxsize=0.39\hsize \epsfbox{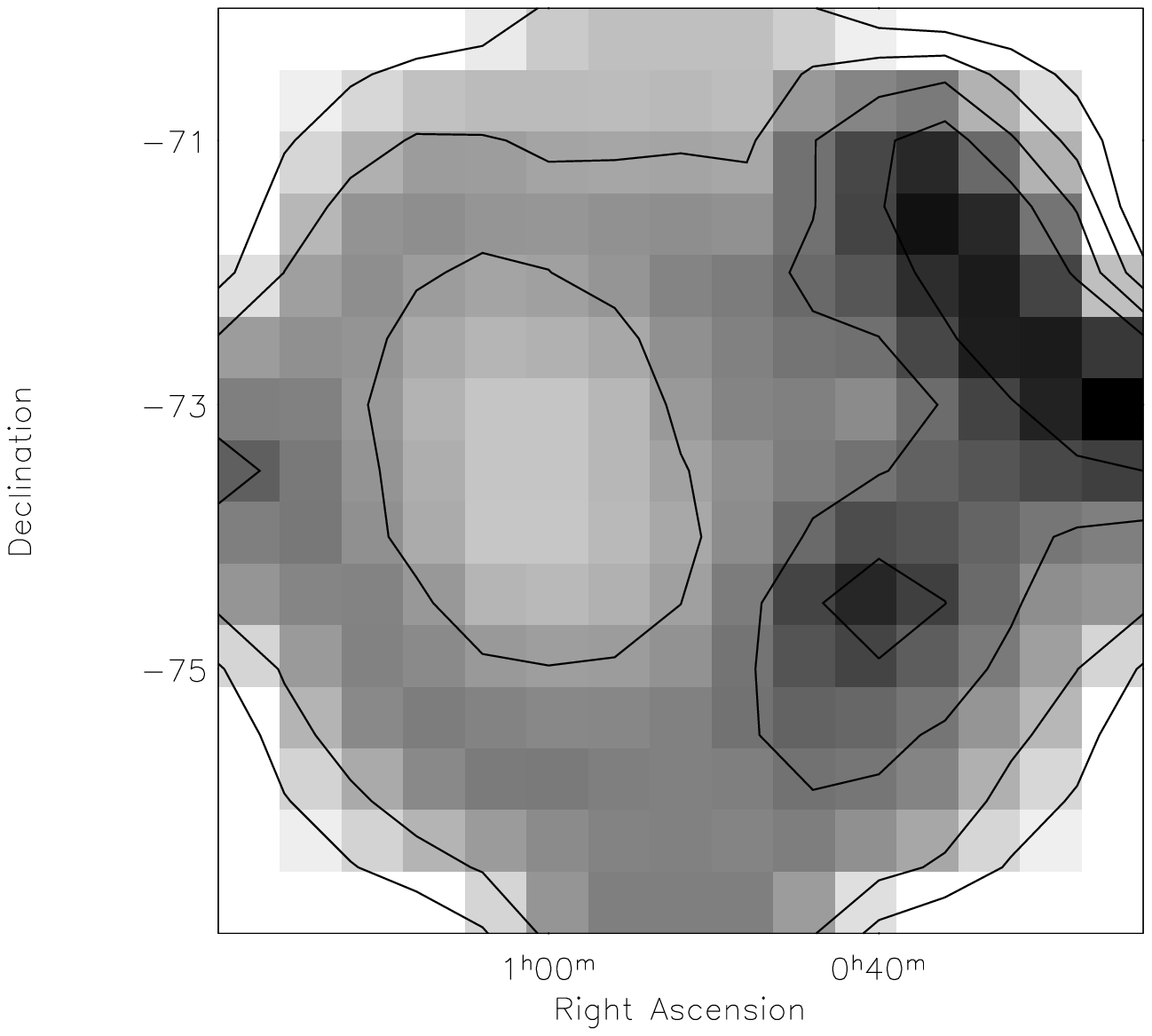}
\hspace{-1.4cm}
\epsfxsize=0.39\hsize \epsfbox{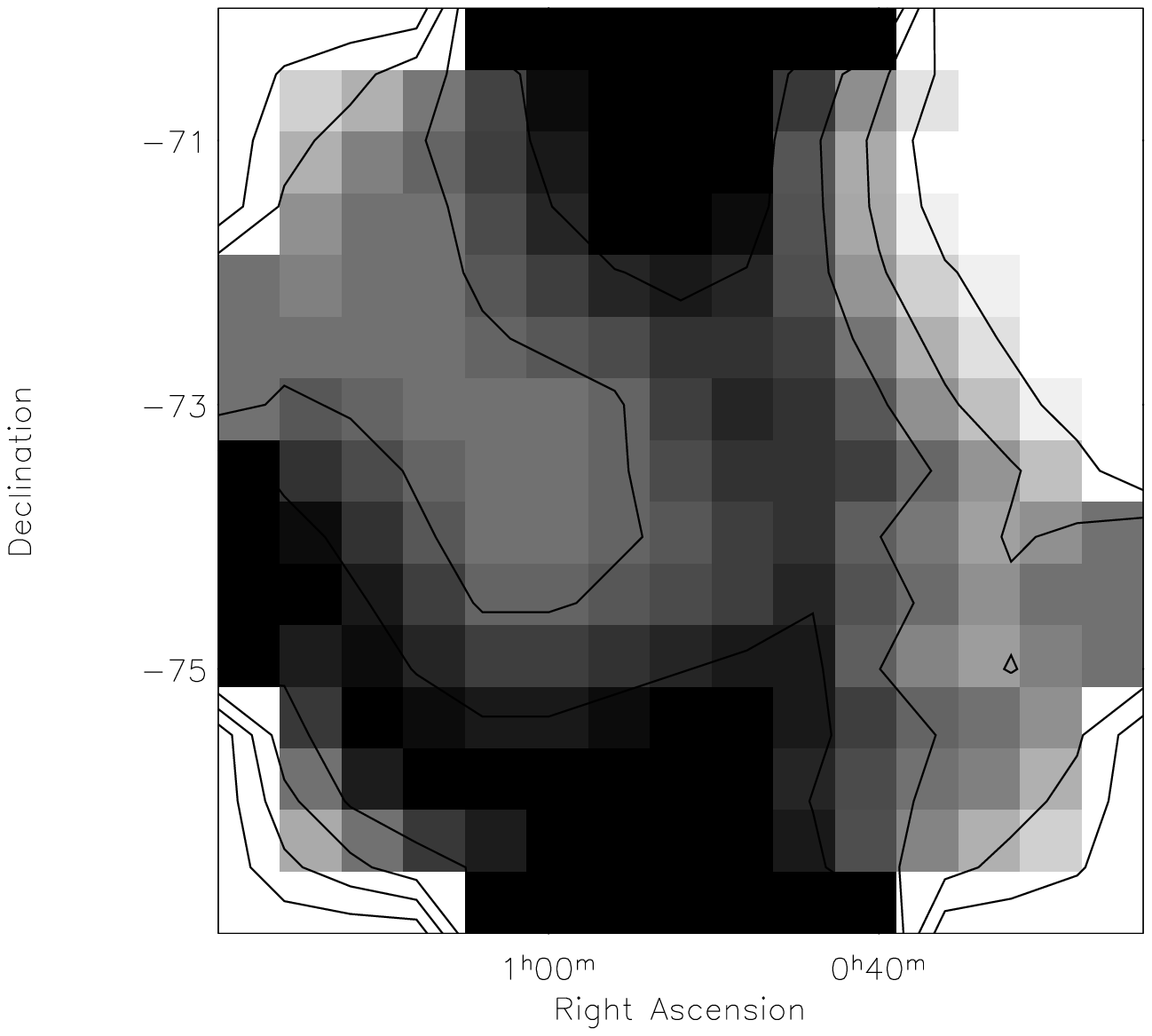}
\hspace{-1.4cm}
\epsfxsize=0.39\hsize \epsfbox{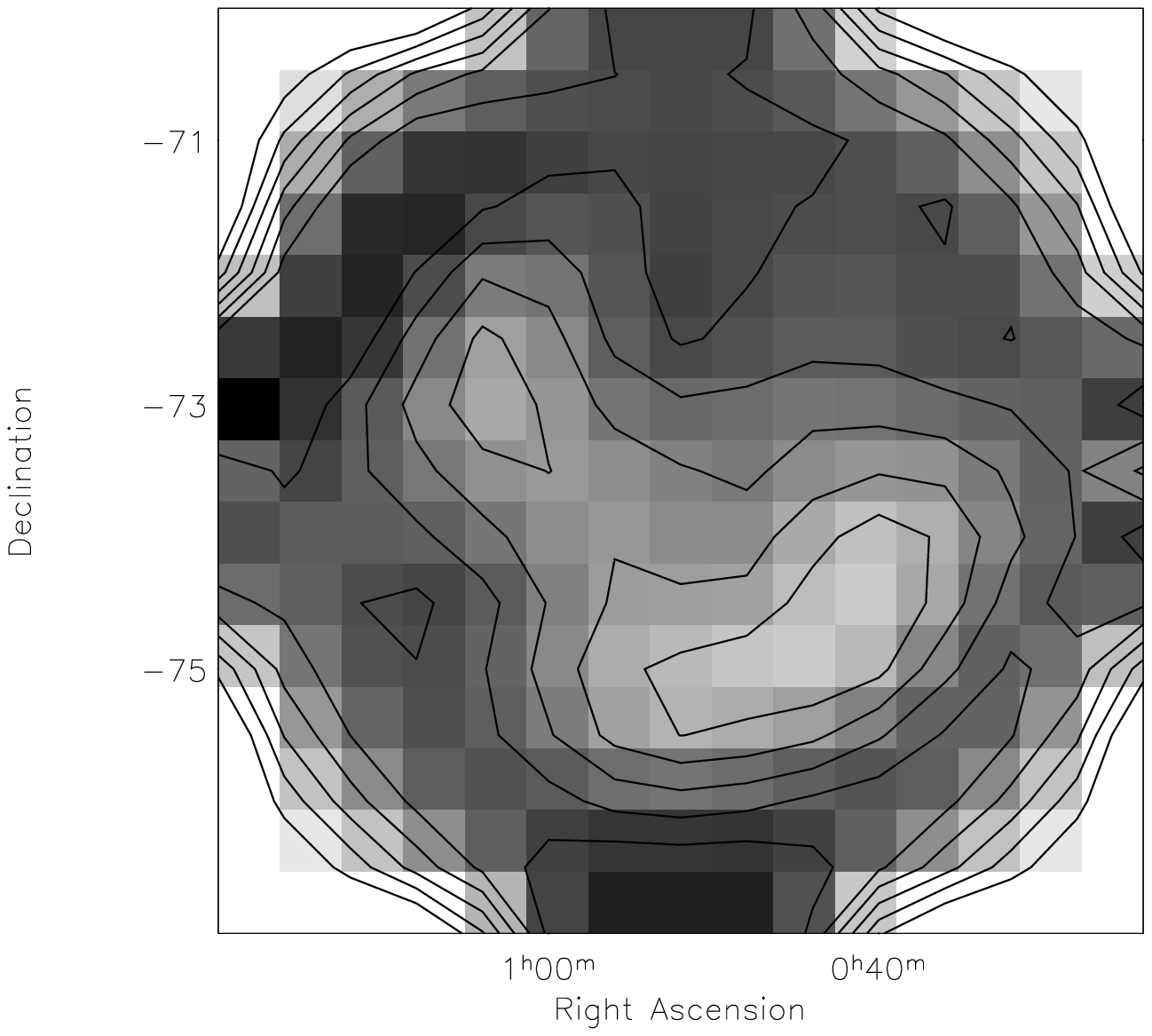} 
\end{minipage}
\begin{minipage}{0.33\hsize}
\caption{{\bf Top row}: distribution of the most probable metallicity
({\bf left} -- contours are at $0.003-0.015$ with a step of $0.002$),
SFR ({\bf middle} -- contours are from $3$ to $10$ with a step of $1$) and the
corresponding probability ({\bf right} -- contours are at $0.7$, $0.8$, $0.9$,
$0.94$ and $0.98$) for C stars. {\bf Bottom row}: the same
distributions for M stars, with the same contour values for both
metallicity and mean age, while the probability has contours at $0.2$,
$0.3$, $0.4$, $0.5$ and $0.6$.} 
\label{lumza}
\end{minipage}
\end{figure*}

\begin{figure}
\hspace{-0.6cm}
\epsfxsize=0.6\hsize \epsfbox{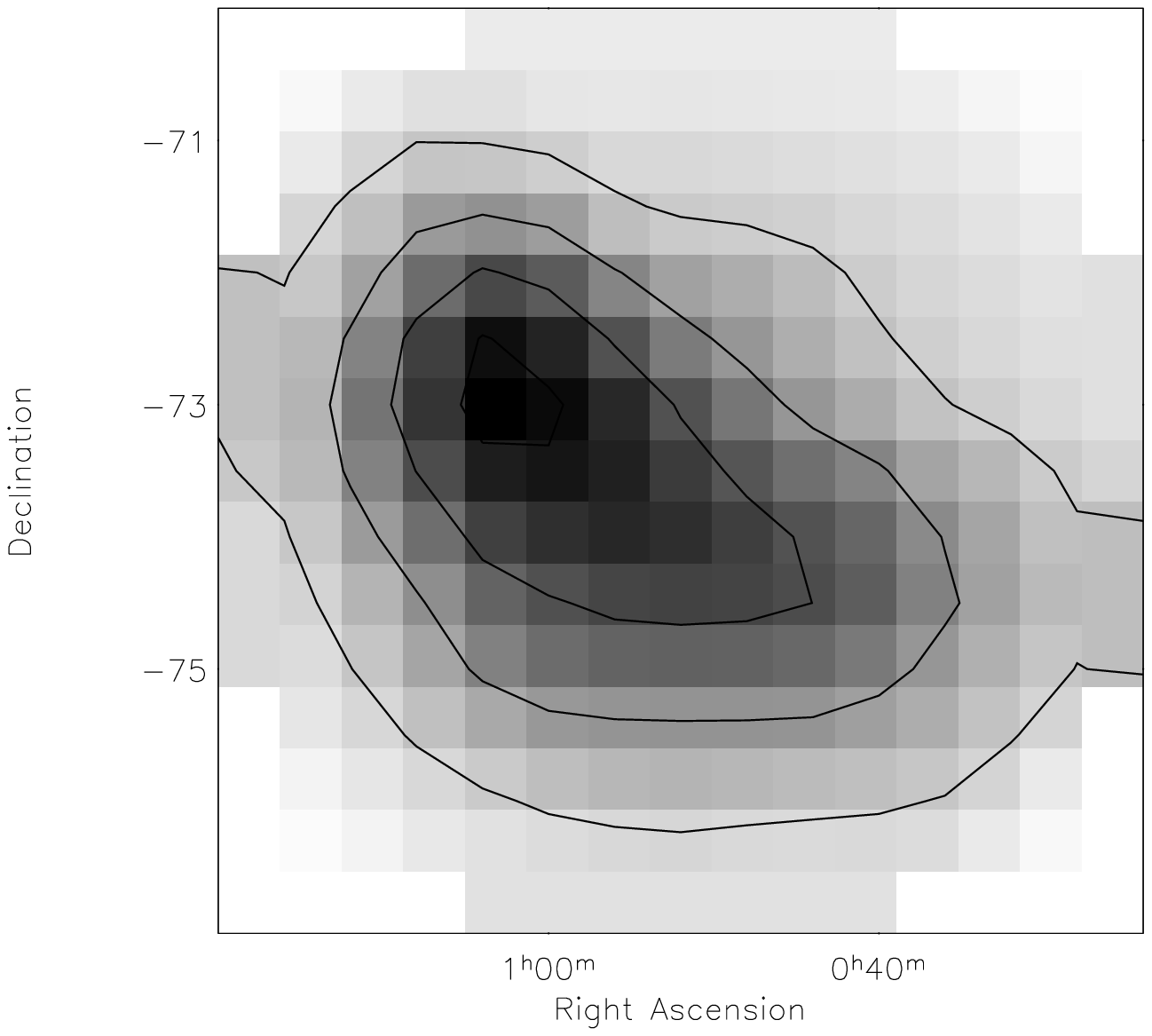}
\hspace{-1.4cm}
\epsfxsize=0.6\hsize \epsfbox{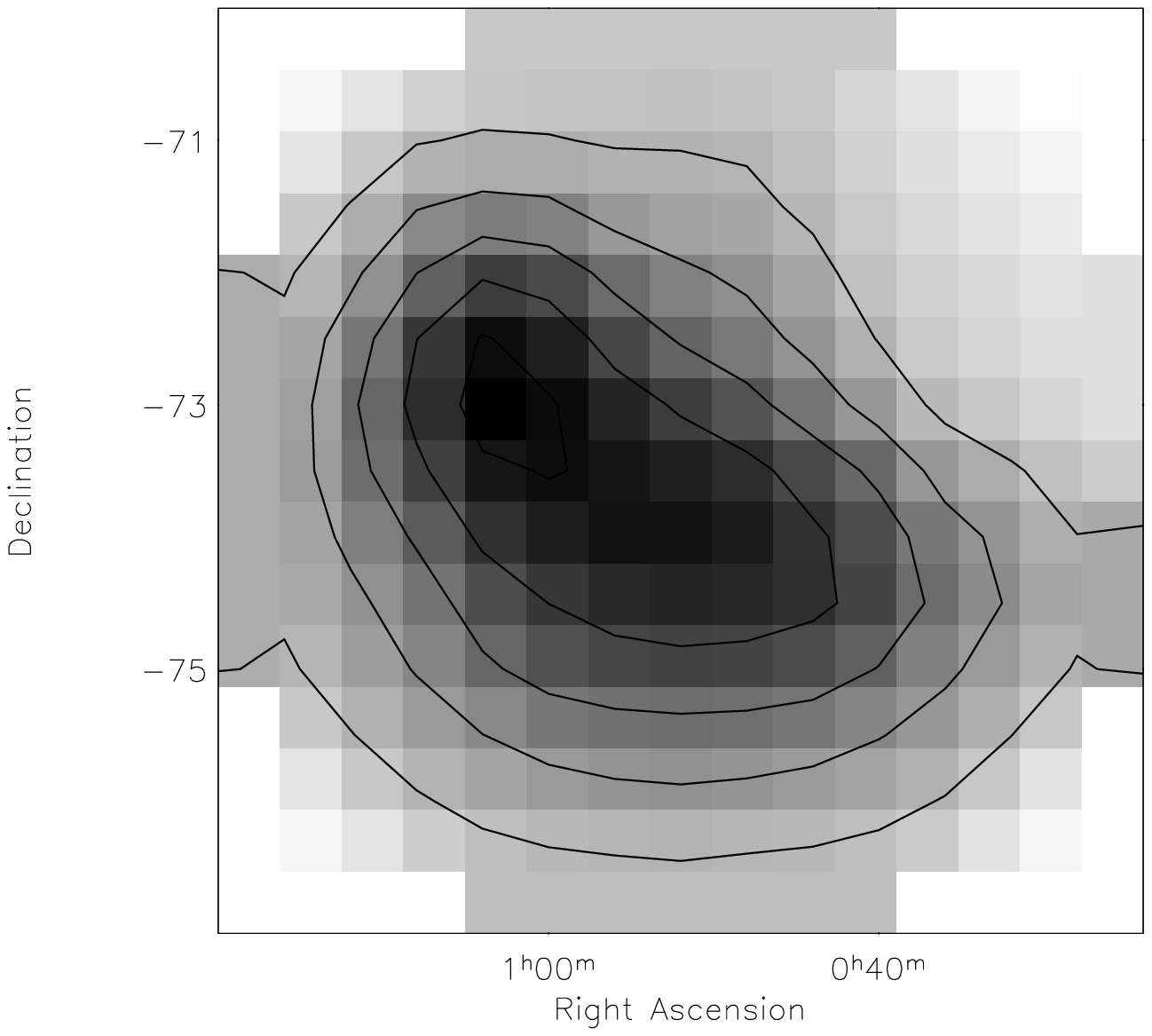}

\hspace{-0.6cm}
\vspace{-0.1cm}
\epsfxsize=0.6\hsize \epsfbox{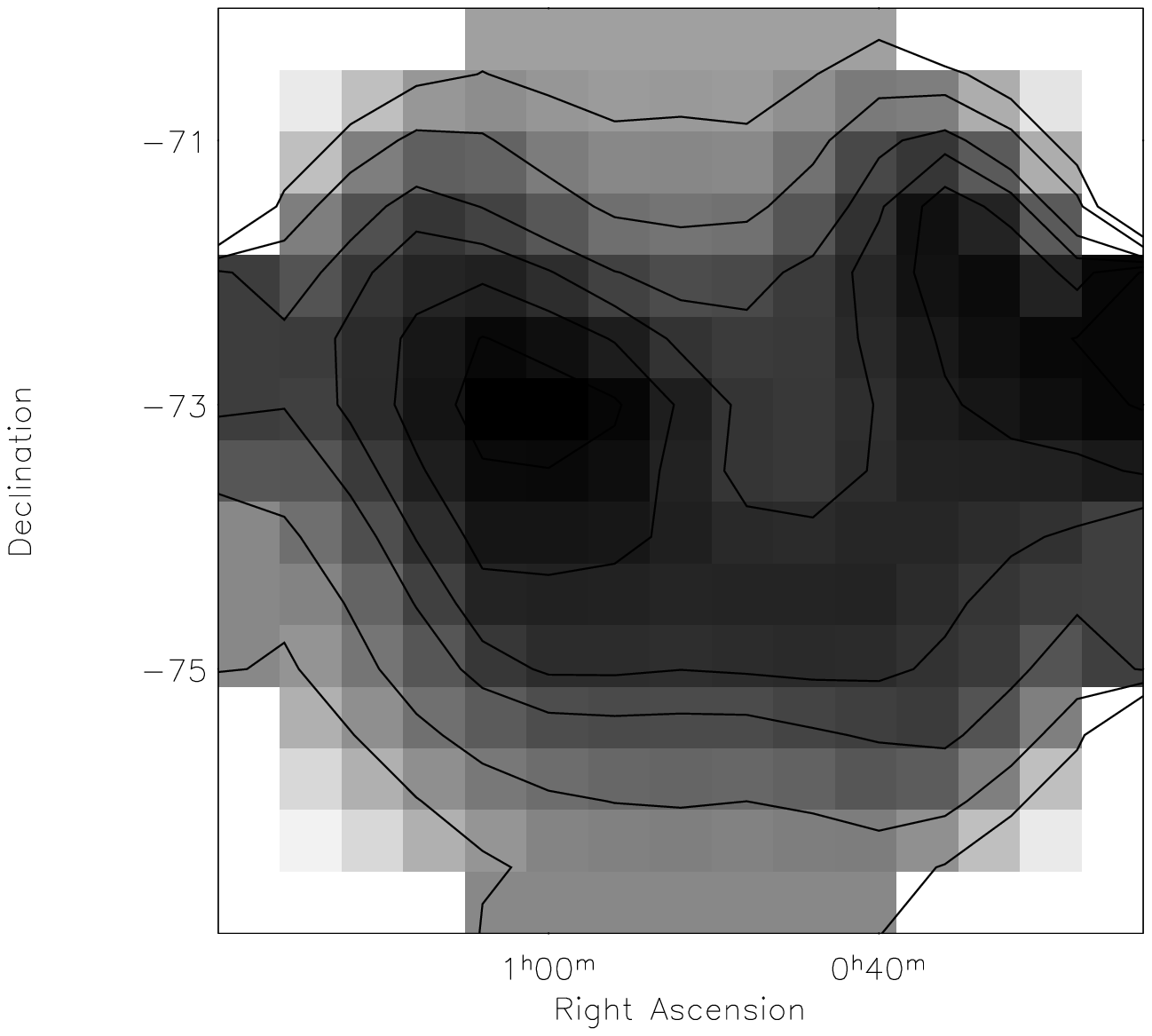}
\hspace{-1.4cm}
\epsfxsize=0.6\hsize \epsfbox{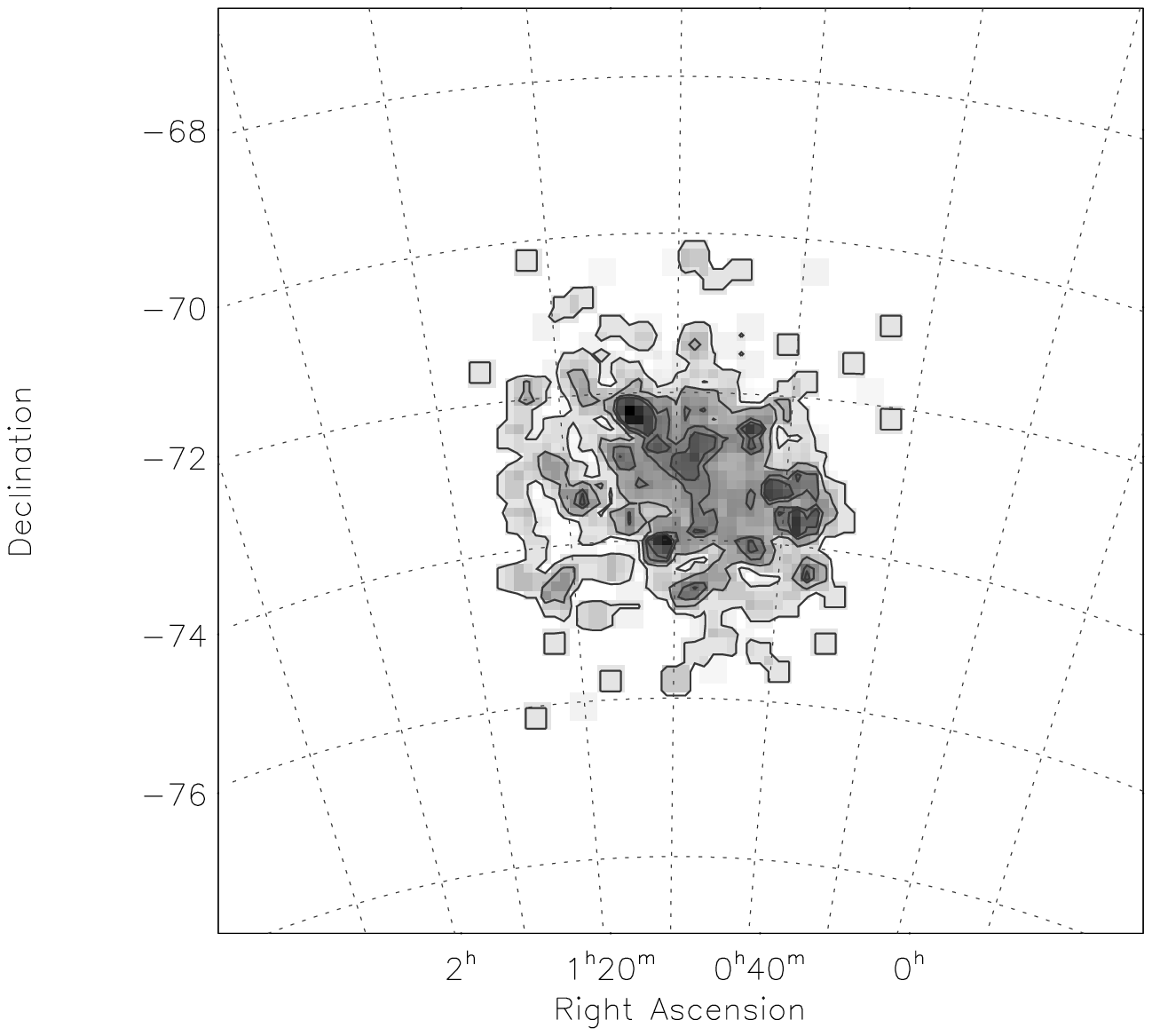} 
\caption{{\bf Top row}: Distribution of the number of C (contours are
at $50$, $100$, $150$, $200$) and M stars (contours are at $100-300$ with a
step of $50$. {\bf
Bottom row}: distribution of the C/M ratio binned as in this work
(contours are at $0.3$, $0.4$, $0.5$, $0.55$, $0.6$, $0.65$) 
and as in Paper I. Note that here because
we need a significant number of stars to fit the LF we do not
reach a resolution as high as in Paper I whilst maintaining the
same global features (see text).}
\label{num}
\end{figure}

In Fig. \ref{lumall} maps of metallicity, separately for C
and M stars across the whole SMC area are shown for each SFR. Contour values
directly correspond to metallicity expressed as values of $Z$. The
distribution of probability is also shown and indicates how
significant a metallicity map is. These maps were created
following three steps: (i) create a grid of $3600$ points in the
plane ($-3:3$, $-3:3$) equally spaced with a step of $0.1$, (ii) 
assign to each point the quantity we are interested (for
example the most probable metallicity) accordingly to which sector
a point belongs to, and (iii) re-bin the distribution of values
in bins equal to $0.4$ (this corresponds to a resolution of $0.16$
deg$^2$) and apply a box car smoothing of width$=2$. 
Finally, we constructed grey scale maps where darker regions
correspond to higher numbers. 
Combining the information in each of the maps in Fig. \ref{lumall}
we show in Fig. \ref{lumza} the metallicity that corresponds to the maximum
probability among the different SFRs as well as the map of SFR and the 
corresponding probability. Note that to each
SFR we have assigned the mean age of all stars formed accordingly to
that particular SFR because this quantity, contrary to
the mean age of AGB stars only, does not depend on metallicity. 
Therefore contour values correspond to the mean age of the
local population. 

Figure \ref{num} shows the distribution of the number of C and
M stars and of their C/M ratio across the SMC with the same binning as
in the other maps. A high resolution version of the distribution of the
C/M ratio as obtained in Paper I is also shown. The
effect of binning in smoothing galaxy features is clear however major
trends are conserved. For example a comparison of both C/M ratio
distributions shows dark regions (i.e. regions with a high C/M
ratio) to the NE and NW of the SMC connected by a southern bridge of
relatively high ratio values. These two darker patches are
recognisable also in the high-resolution map as well as another dark patch
to the 
S of the galaxy centre. The outer SMC has a low C/M ratio in both
maps. High resolution maps of all AGB stars (C
plus M stars) can be found in Cioni et al. (\cite{cmor}).

The distribution of C stars is described by a broken ring-like structure of
high metallicity corresponding to an average $Z=0.006$ with higher
values to the N. The centre and the wing region correspond to an
average metallicity as low as $Z=0.003$ or lower.
The mean age of the stellar population is $8-10$ Gyr in the wing
(SE) and to the W. Younger ($<7$ Gyr) populations are
present in the NE and perhaps in the centre of the SMC while
the outermost regions are perhaps influenced by a low number
statistics. These suggestions
about the distribution of metallicity and mean age of the underlying stellar
population deduced from C stars correspond to a $90$\% confidence
level within most of the galaxy. This is true also for the individual
maps shown in Fig. \ref{lumall} (first \& second row). It is
interesting to note that a ring-like metal-rich structure is present
for any SFR considered. However the specific location of the peak 
value of Z moves clockwise from the SE to the NW as a
function of increasing look-back time.
A much lower confidence level is associated to the maps obtained from
M stars in particular along the bar of the galaxy. 
These indicate that most of the population has an average
metallicity corresponding to $Z=0.006$ with higher values to the
NW ($Z=0.009$ or higher) and lower values E of the galaxy
centre ($Z=0.003$) and in the outermost regions. This
metallicity distribution corresponds to an older ($>6$ Gyr) stellar
population in the SE and to the N, while in the NE, NW and 
SW the stellar population appears younger.
Although results obtained from M stars are less reliable than those
obtained from C stars they both suggest that the NW of the
galaxy has a higher metallicity compared to the rest of it, and that
there is an old population in the wing region compared to a
younger population in the centre of the galaxy.

\section{Discussion}
\label{dis}
\subsection{How well does the C/M ratio alone traces metallicity?}
In Paper I the distribution of the C/M ratio has been interpreted
purely as a tracer of metallicity 
concluding that in the SMC, probably because of its complex structure, there
is no clear trend but patches of varying metallicity, perhaps a
metal rich outer ring and metal poor regions in the centre of the
galaxy. In Paper II we have shown that the C/M ratio depends
also on the SFR and we have created maps of metallicity, separately 
for C and M stars for the LMC, that account for this effect. Here
we have modified our algorithm to produce similar maps for the SMC. 
Contrary to the LMC analysis the distribution of the C/M
ratio across the SMC (Fig. \ref{num}) and the distributions of
metallicity shown in Fig. \ref{lumza} have little in common. 
The region with high C/M values (i.e. a low metallicity) E of the
centre corresponds to a metal poor (Z $\approx 0.003$) 
$6-7$ Gyr old population. On the other hand a region with a similarly 
high C/M ratio to the NW corresponds to $Z=0.009$ with a
controversial mean age, either very young from M stars or very old
from C stars, but the confidence level in this area is poor 
($<70$\%). The bridge connecting the two regions of high C/M ratio are
low in metals also from the interpretation of the LF of C stars, while
the LF of M stars in this area suggests a higher metallicity and a  
population $5-6$ Gyr old. To the N of the galaxy  the C/M ratio has a low 
value that corresponds to a metal rich ($Z=0.009$) population also of
intermediate-age ($\approx 7$ Gyr) in the maps obtained from C stars
while M stars maps indicate a lower metallicity ($Z=0.006$) but a
similar mean age. 
We conclude that the C/M ratio is a good tracer of metallicity only
where the underlying stellar population is of intermediate-age ($5-7$ Gyr).
Considering the difference between the low and high metallicity
contour in Fig. \ref{lumza}, converted into [M/H] assuming
$Z_\odot=0.019$, we obtain a spread of at least $0.5$ dex. Perhaps when
we will understand the orientation of the SMC, including its
extension along the line of sight, we may reach a different conclusion. 

\subsection{Comparison with the literature}
Combining the information about the distribution of the mean age of
the stellar population across the SMC 
obtained interpreting the LF of both C and M stars we conclude that the
overall SMC stellar population is of intermediate age (in agreement 
with Hardy \& Durand
\cite{hardur}) and that the outermost stellar population, including the
wing region, appears older. 
In fact here we do not detect a younger component as
suggested by Gardiner \& Hatzidimitriou (\cite{garhat}) although 
results overlap partly only in the inner wing region (see their
Fig. 5 and Fig. 7). However, we cannot exclude that part of the stellar
population towards the direction of the LMC (NE) is as young
as a few Gyr, on average younger than the overall SMC population. 
In the region observed by Dolphin et al. (\cite{dol}) using HST we confirm a
mean-age of about $8$ Gyr. 

Even more interesting is the comparison of our results with those of Harris \& 
Zaritsky (\cite{harr}), because they also provide maps of the star formation 
history and mean metallicity of the SMC populations, maps which are a factor 
of four more detailed on both the spatial and age distributions 
($0.04$ deg$^2$ versus $0.16$ deg$^2$). Their results were 
derived from an extensive analysis of optical photometric data, which better 
sample stars that we have eliminated from our analysis, namely dwarfs, 
subgiants and giants of spectral type earlier than M. AGB stars, which are 
the only stars considered by us, are likely to play a negligible role in 
their study. 
Harris \& Zaritsky (\cite{harr}) suggest that the main body of the
SMC is surrounded by a ring of moderately metal rich stars ($Z=0.008$)
about $2.5$ Gyr old (see their Fig. 6). This structure corresponds to
the broken ring-like structure outlined by the distribution of 
metallicity obtained from the LF of C stars (Fig. \ref{lumza})
eventhough the ring is not connected and indicates a 
more metal poor population towards its southern part. The mean age that we
derive for this structure is also not homogeneous and seems
to indicate the presence of older stars in the W and in the
SE of the broken ring compared to other locations. 
In addition Harris \& Zaritsky (\cite{harr}) derive that the
population in the centre of the galaxy spans a 
large range of ages: from $10$ Gyr old, to the NE of it, to
$1$ Gyr and as young as 
$40$ Myr old in the SW of it. This explains why we attribute
to the stars in the central region an average age of $5-7$ Gyr  
although we conclude that this region is relatively metal poor 
and, accordingly to a gradual and monotonic increase of the age-metallicity 
relation, is supported by a predominantly old stellar population. 

It is common practice to attribute to the SMC an overall metallicity
corresponding to $Z=0.004$. The results discussed in this paper, as
well as those obtained by Harris \& Zaritsky (\cite{harr}), prove that
it is not sufficient to describe the whole galaxy with a unique
metallicity. The spread as a function of position is not negligible as
well as the age of the stellar population that contributes to it.
Thus, there seems to be a good overall agreement between the main results 
reached by Harris \& Zaritsky (\cite{harr}) and ours, at least regarding 
those features discussed above. This is certainly 
reassuring and supports the validity of our approach based on AGB stars. 
There are however some differences between the two methods. For instance, 
Harris \& Zaritsky provide SFR and metallicity maps with a finer resolution 
which is unreachable using our method because AGB stars are simply too few 
compared to ``normal stars'' -- mainly dwarfs and subgiants -- 
used by them. Moreover, by relying on star counts of normal stars, Harris 
\& Zaritsky (\cite{harr}) provide star formation histories with an 
age-resolution corresponding to age bins of just 
$\Delta\log({\rm age})=0.2$ dex wide; 
such age details are unlikely to be reached using AGB stars only, since 
effects such as long-period variability and model uncertainties still 
hamper the use of AGB stars as accurate stellar clocks.

However, the use of AGB stars provides interesting possibilities, either 
unreachable by, or complementary to, those using optical data. First, it is 
easier to sample the complete AGB population of a Local Group galaxy in the 
near-infrared than obtaining its optical photometry complete down to the 
oldest turn-off across the whole galaxy; therefore, the global analysis of 
the AGB population might be presently applied to a larger number of 
galaxies. Second, in using the near-infrared photometry, we reduce 
considerably the complications induced by differential reddening. In fact, 
Harris \& Zaritsky (\cite{harr})  were able to apply suitable reddening 
corrections to their data just because they had photometry in four 
pass-bands ($UBVI$) resulting from a remarkable observational effort. 
Even though, in the course of their work they found a previously 
unsuspected variation of extinction with the stellar population 
(Zaritsky \cite{zar99}, Zaritsky et al. \cite{zar04}). It is clear that 
reddening constitutes a major complication in the analysis of stars 
formation histories using optical data, and that this is not the case 
when using AGB stars. Finally another advantage is represented by crowding,
$K_s$-band images of most galaxies that can be resolved into stars are 
not affected by crowding compared to their $I$-band counterpart. Harris \& 
Zaritsky (\cite{harr}) estimated photometric errors via a linear correlation 
with the stellar density, which is certainly a major component of the error 
but not the only one.

\section{Summary and Conclusions}
\label{fin}
In this paper we compare the observed $K_{\mathrm s}$ magnitude
distribution of C and M stars within sectors covering entirely
the SMC, of a suitable size to provide a statistical
significant sample, with theoretical distributions. These are
derived employing suitable stellar evolutionary tracks for stars of
different mass and in particular using an improved version of
Marigo et al. (\cite{ma03}) models for TP-AGB stars.
This allows us to well describe the behaviour of
both C and M type stars as a function of $K_{\mathrm s}$ magnitude.
Details about the models are given in Paper II of this series as well as 
a similar study across the LMC.
The quality of the comparison has been quantified using a $\chi^2$ test.

Surface maps indicating the most probable metallicity and mean age
distribution have been obtained varying one or
both parameters at the same time. Results have been
compared with the information available from previous studies.
With respect to the conclusions reached in Paper I, the C/M ratio is
influenced by the age of the population where it is
calculated, however, it remains a good tracer of the metallicity only if
the population is of intermediate-age ($5-7$ Gyr old). 
Note that here we refer to the mean age of the
population in a given spatial location and not to the precise age at which
that particular population began forming stars. 

Our best fit models indicate that 
stars in the SMC are overall older and metal poorer than stars in the
LMC. The average SMC population is $7-9$ Gyr old with older stars in
the outermost regions except perhaps towards the LMC where younger
stars may be present. Note that a close passage of the LMC occurred
about $6.5$ Gyr ago and may have triggered star formation in the SMC,
many clusters were also formed at this epoch (Piatti et al. \cite{pia}). 
The metallicity along a ring-like structure
surrounding the main body of the galaxy varies from $Z=0.003$ to $Z=0.009$
with peak values towards the NW while
towards the S, comprising the centre of the galaxy, the
metallicity is probably lower. However, since TP-AGB models are not
perfect, and since we are testing just a family (the exponential ones)
among many possible SFR histories, these values of age and metallicity
may be affected by systematic errors, and are likely to change as we
adopt improved TP-AGB models, and test other possible SFR functions. 
Instead, we consider to be more significant the detected variations of
the values of these parameters (i.e. age and metallicity) across the SMC. 
Information about the structure of the
SMC as well as about its dynamical history are also desirable to constraint its
star formation history distribution.

\end{document}